\documentclass[floats,floatfix,showpacs,amsmath,amssymb,floatfix,prd,twocolumn,superscriptaddress,nofootinbib,nolongbibliography,reprint]{revtex4-2}

\def\apjl{{Astrophys.~J.~Lett.}}
\def\apjs{{Astrophys.~J.~Supp.}}
\def\ao{{Appl.~Opt.}}

\def\aap{{Astron.~Astrophys.}}

\def\prd{{Phys.~Rev.~D}}

\def\prl{{Phys.~Rev.~Lett.}}

\def\nat{{Nature}}

\def\lrr{{Living Rev. Relativ.}}

\usepackage{microtype}
\usepackage[utf8]{inputenc} 
\usepackage{graphicx}
\usepackage{pifont}% Checkmark and xmark in tables
\newcommand{\cmark}{\ding{51}}%
\newcommand{\xmark}{\ding{55}}%
\usepackage{dcolumn}% Align table columns on decimal point
\usepackage{bm}% bold math
\usepackage{lineno}
\usepackage[usenames, dvipsnames, table]{xcolor}
\definecolor{linkcolor}{rgb}{0.0,0.3,0.5}
\usepackage[colorlinks = true,
            linkcolor = linkcolor,
            urlcolor  = linkcolor,
            citecolor = linkcolor,
            anchorcolor = linkcolor]{hyperref}  
\usepackage{array,afterpage,placeins}

\newcommand\orcidlink[1]{\href{https://orcid.org/#1}{\includegraphics[scale=0.006]{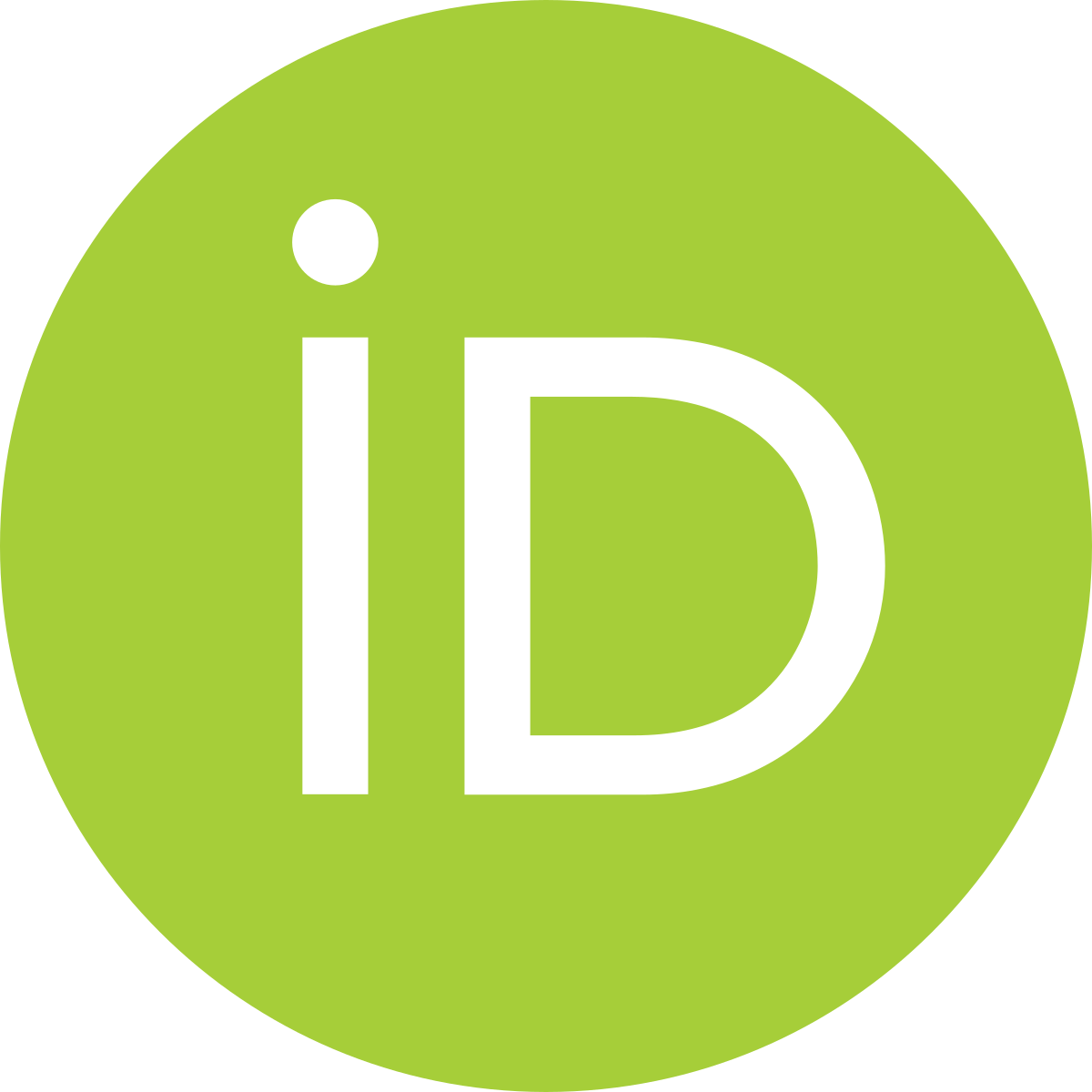}}}

\begin{document}
\title{Glitch systematics on the observation of massive black-hole binaries with LISA}

\newcommand{\bham}{\affiliation{Institute for Gravitational Wave Astronomy \& School of Physics and Astronomy, University of Birmingham, Birmingham, B15 2TT, UK}}
\newcommand{\milan}{\affiliation{Dipartimento di Fisica ``G. Occhialini'', Universit\`a degli Studi di Milano-Bicocca, Piazza della Scienza 3, 20126 Milano, Italy} \affiliation{INFN, Sezione di Milano-Bicocca, Piazza della Scienza 3, 20126 Milano, Italy}}
\newcommand{\unitn}{\affiliation{Department of Physics, University of Trento, and Trento Institute for Fundamental Physics and Applications, INFN,  38123 Povo, Trento, Italy}
}
%Authors
\author{Alice Spadaro~\orcidlink{0009-0000-8104-6171}}
\email{a.spadaro3@campus.unimib.it}
\milan
\author{Riccardo Buscicchio~\orcidlink{0000-0002-7387-6754}}
\email{riccardo.buscicchio@unimib.it}
\milan
\bham
\author{Daniele Vetrugno~\orcidlink{0000-0003-0937-1468}}
\unitn
\author{Antoine Klein~\orcidlink{0000-0001-5438-9152}}
\bham
\author{\\Davide Gerosa~\orcidlink{0000-0002-0933-3579}}
\milan \bham
\author{Stefano~Vitale~\orcidlink{0000-0002-2427-8918}}
\unitn
\author{Rita Dolesi~\orcidlink{0000-0002-4191-7558}}
\unitn
\author{William Joseph Weber~\orcidlink{0000-0003-1536-2410}}
\unitn
\author{Monica Colpi~\orcidlink{0000-0002-3370-6152}}
\milan
\date{\today}
%This file is machine generated. Do *not* edit it manually if you want reproducibility.
%Hashed ..//run1/config_inj.yaml: 34eb18acc78c100178dd1e49917e945d
%Hashed ..//run1/config_pe.yaml: 77a9696535544766fdd2b689250dccc6
%Hashed ..//run1/physical_posterior.dat: a120161dd38e4eb1a2727341fc6a7fe3
\newcommand{\RunOneBHBOneDimensionlessSpinOneZero}{\ensuremath{0.4^{+0.1}_{-0.1}}}
\newcommand{\RunOneBHBOneDimensionlessSpinTwoZero}{\ensuremath{0.3^{+0.6}_{-1.0}}}
\newcommand{\RunOneBHBOneEclipticLatitudeZero}{\ensuremath{0.3^{+0.6}_{-0.1}}}
\newcommand{\RunOneBHBOneEclipticLongitudeZero}{\ensuremath{2.0^{+0.2}_{-0.1}}}
\newcommand{\RunOneBHBOneInclinationZero}{\ensuremath{0.8^{+0.3}_{-0.6}}}
\newcommand{\RunOneBHBOneInitialOrbitalPhaseZero}{\ensuremath{1.6^{+1.5}_{-1.3}}}
\newcommand{\RunOneBHBOneLuminosityDistanceZero}{\ensuremath{44^{+15}_{-14}}}
\newcommand{\RunOneBHBOneMergerTimeOrInitialOrbitalFrequencyZero}{\ensuremath{30.01^{+0.09}_{-0.08}}}
\newcommand{\RunOneBHBOnePolarizationZero}{\ensuremath{1.6^{+1.3}_{-1.4}}}
\newcommand{\RunOneBHBOneRedshiftedMassOneZero}{\ensuremath{4.5^{+0.2}_{-0.2}}}
\newcommand{\RunOneBHBOneRedshiftedMassTwoZero}{\ensuremath{1.5^{+0.3}_{-0.3}}}
\newcommand{\RunOneInjBHBOneDimensionlessSpinOneZero}{\ensuremath{0.4}}
\newcommand{\RunOneInjBHBOneDimensionlessSpinTwoZero}{\ensuremath{0.3}}
\newcommand{\RunOneInjBHBOneEclipticLatitudeZero}{\ensuremath{0.30}}
\newcommand{\RunOneInjBHBOneEclipticLongitudeZero}{\ensuremath{2.0}}
\newcommand{\RunOneInjBHBOneInclinationZero}{\ensuremath{0.64}}
\newcommand{\RunOneInjBHBOneInitialOrbitalPhaseZero}{\ensuremath{1.0}}
\newcommand{\RunOneInjBHBOneLuminosityDistanceZero}{\ensuremath{47.6}}
\newcommand{\RunOneInjBHBOneMergerTimeOrInitialOrbitalFrequencyZero}{\ensuremath{30.0}}
\newcommand{\RunOneInjBHBOnePolarizationZero}{\ensuremath{1.7}}
\newcommand{\RunOneInjBHBOneRedshiftedMassOneZero}{\ensuremath{4.5}}
\newcommand{\RunOneInjBHBOneRedshiftedMassTwoZero}{\ensuremath{1.5}}
%File created on 29-05-2023 17:22:37

%This file is machine generated. Do *not* edit it manually if you want reproducibility.
%Hashed ..//run2/config_inj.yaml: 927217126b37fb69b2b09da0ae3d4ae7
%Hashed ..//run2/config_pe.yaml: 491d22db1eba87a1df980ecd2fe9136f
%Hashed ..//run2/physical_posterior.dat: 6149530d48ea8b25e95b5b48c33e5a47
\newcommand{\RunTwoBHBOneDimensionlessSpinOneZero}{\ensuremath{0.4^{+0.1}_{-0.1}}}
\newcommand{\RunTwoBHBOneDimensionlessSpinTwoZero}{\ensuremath{0.3^{+0.6}_{-1.0}}}
\newcommand{\RunTwoBHBOneEclipticLatitudeZero}{\ensuremath{0.3^{+0.5}_{-0.1}}}
\newcommand{\RunTwoBHBOneEclipticLongitudeZero}{\ensuremath{1.99^{+0.25}_{-0.09}}}
\newcommand{\RunTwoBHBOneInclinationZero}{\ensuremath{0.8^{+0.3}_{-0.6}}}
\newcommand{\RunTwoBHBOneInitialOrbitalPhaseZero}{\ensuremath{1.5^{+1.5}_{-1.3}}}
\newcommand{\RunTwoBHBOneLuminosityDistanceZero}{\ensuremath{44^{+15}_{-14}}}
\newcommand{\RunTwoBHBOneMergerTimeOrInitialOrbitalFrequencyZero}{\ensuremath{30.01^{+0.09}_{-0.08}}}
\newcommand{\RunTwoBHBOnePolarizationZero}{\ensuremath{1.6^{+1.3}_{-1.4}}}
\newcommand{\RunTwoBHBOneRedshiftedMassOneZero}{\ensuremath{4.5^{+0.2}_{-0.2}}}
\newcommand{\RunTwoBHBOneRedshiftedMassTwoZero}{\ensuremath{1.5^{+0.3}_{-0.3}}}
\newcommand{\RunTwoInjBHBOneDimensionlessSpinOneZero}{\ensuremath{0.4}}
\newcommand{\RunTwoInjBHBOneDimensionlessSpinTwoZero}{\ensuremath{0.3}}
\newcommand{\RunTwoInjBHBOneEclipticLatitudeZero}{\ensuremath{0.30469265401539747}}
\newcommand{\RunTwoInjBHBOneEclipticLongitudeZero}{\ensuremath{2.0}}
\newcommand{\RunTwoInjBHBOneInclinationZero}{\ensuremath{0.6435011087932843}}
\newcommand{\RunTwoInjBHBOneInitialOrbitalPhaseZero}{\ensuremath{1.0}}
\newcommand{\RunTwoInjBHBOneLuminosityDistanceZero}{\ensuremath{47.6}}
\newcommand{\RunTwoInjBHBOneMergerTimeOrInitialOrbitalFrequencyZero}{\ensuremath{30.0}}
\newcommand{\RunTwoInjBHBOnePolarizationZero}{\ensuremath{1.7}}
\newcommand{\RunTwoInjBHBOneRedshiftedMassOneZero}{\ensuremath{4.5}}
\newcommand{\RunTwoInjBHBOneRedshiftedMassTwoZero}{\ensuremath{1.5}}
%File created on 29-05-2023 13:55:23

%File created on 07-05-2023 01:40:17
%This file is machine generated. Do *not* edit it manually if you want reproducibility.
%Hashed ../results/final_runs//run3/config_inj.yaml: bde37562f2f2938772c94d23fe24d675
%Hashed ../results/final_runs//run3/config_pe.yaml: bde37562f2f2938772c94d23fe24d675
%Hashed ../results/final_runs//run3/physical_posterior.dat: b7cd00157f977c663ae795b71a506ec4
\newcommand{\RunThreeSLOneAmplitudeDisplacementZero}{\ensuremath{0.095^{+0.004}_{-0.003}}}
\newcommand{\RunThreeSLOneBetaZero}{\ensuremath{40.53^{+10.78}_{-9.86}}}
\newcommand{\RunThreeSLOneInitialTimeZero}{\ensuremath{3559^{+17}_{-18}}}
\newcommand{\RunThreeInjSLOneAmplitudeDisplacementZero}{\ensuremath{0.0948}}
\newcommand{\RunThreeInjSLOneBetaZero}{\ensuremath{40.0}}
\newcommand{\RunThreeInjSLOneInitialTimeZero}{\ensuremath{3560.0}}
%File created on 07-05-2023 01:34:06
%This file is machine generated. Do *not* edit it manually if you want reproducibility.
%Hashed ../results/final_runs//run4/config_inj.yaml: 86a9bff54dc121583bc270140b80637d
%Hashed ../results/final_runs//run4/config_pe.yaml: 86a9bff54dc121583bc270140b80637d
%Hashed ../results/final_runs//run4/physical_posterior.dat: 3a3111def1859e62ebc3eb6e9d8995e2
\newcommand{\RunFourLPFOneDeltavZero}{\ensuremath{3.01^{+0.10}_{-0.09}}}
\newcommand{\RunFourLPFOneInitialTimeZero}{\ensuremath{1800^{+22}_{-21}}}
\newcommand{\RunFourLPFOneTauFallZero}{\ensuremath{1722^{+325}_{-270}}}
\newcommand{\RunFourLPFOneTauRiseZero}{\ensuremath{1569^{+291}_{-246}}}
\newcommand{\RunFourInjLPFOneDeltavZero}{\ensuremath{3.0}}
\newcommand{\RunFourInjLPFOneInitialTimeZero}{\ensuremath{1800.0}}
\newcommand{\RunFourInjLPFOneTauFallZero}{\ensuremath{1800.0}}
\newcommand{\RunFourInjLPFOneTauRiseZero}{\ensuremath{1500.0}}
%File created on 07-05-2023 01:34:20
%This file is machine generated. Do *not* edit it manually if you want reproducibility.
%Hashed ../results/final_runs//run5/config_inj.yaml: 4253682d3482c55cb4d14889c1eb177e
%Hashed ../results/final_runs//run5/config_pe.yaml: 4d547d4d7c817ba7663e8e08f18c4a9e
%Hashed ../results/final_runs//run5/physical_posterior.dat: 62c001315c53f36065b97a73813f15ae
\newcommand{\RunFiveSLOneAmplitudeDisplacementZero}{\ensuremath{200^{+27}_{-24}}}
\newcommand{\RunFiveSLOneBetaZero}{\ensuremath{5.2^{+3.6}_{-3.4}}}
\newcommand{\RunFiveSLOneInitialTimeZero}{\ensuremath{3559.7^{+6.1}_{-6.1}}}
\newcommand{\RunFiveInjSLOneAmplitudeDisplacementZero}{\ensuremath{200.0}}
\newcommand{\RunFiveInjSLOneBetaZero}{\ensuremath{5.0}}
\newcommand{\RunFiveInjSLOneInitialTimeZero}{\ensuremath{3560.0}}

%This file is machine generated. Do *not* edit it manually if you want reproducibility.
%Hashed ..//run6/config_inj.yaml: aacec94b12e58101646c4839aeffac43
%Hashed ..//run6/config_pe.yaml: d759171e17c1b56d6ded10ee93507e58
%Hashed ..//run6/physical_posterior.dat: 794737707ac7bd31f9d524983dd72763
\newcommand{\RunSixBHBOneDimensionlessSpinOneZero}{\ensuremath{-0.308^{+0.005}_{-0.005}}}
\newcommand{\RunSixBHBOneDimensionlessSpinTwoZero}{\ensuremath{-0.51^{+0.09}_{-0.09}}}
\newcommand{\RunSixBHBOneEclipticLatitudeZero}{\ensuremath{0.150^{+0.005}_{-0.005}}}
\newcommand{\RunSixBHBOneEclipticLongitudeZero}{\ensuremath{1.792^{+0.005}_{-0.005}}}
\newcommand{\RunSixBHBOneInclinationZero}{\ensuremath{0.04^{+0.05}_{-0.03}}}
\newcommand{\RunSixBHBOneInitialOrbitalPhaseZero}{\ensuremath{1.6^{+1.4}_{-1.4}}}
\newcommand{\RunSixBHBOneLuminosityDistanceZero}{\ensuremath{10.010^{+0.032}_{-0.009}}}
\newcommand{\RunSixBHBOneMergerTimeOrInitialOrbitalFrequencyZero}{\ensuremath{29.447^{+0.007}_{-0.002}}}
\newcommand{\RunSixBHBOnePolarizationZero}{\ensuremath{1.6^{+1.4}_{-1.4}}}
\newcommand{\RunSixBHBOneRedshiftedMassOneZero}{\ensuremath{4.25^{+0.01}_{-0.01}}}
\newcommand{\RunSixBHBOneRedshiftedMassTwoZero}{\ensuremath{0.610^{+0.007}_{-0.007}}}
\newcommand{\RunSixInjBHBOneDimensionlessSpinOneZero}{\ensuremath{0.4}}
\newcommand{\RunSixInjBHBOneDimensionlessSpinTwoZero}{\ensuremath{0.3}}
\newcommand{\RunSixInjBHBOneEclipticLatitudeZero}{\ensuremath{0.30469265401539747}}
\newcommand{\RunSixInjBHBOneEclipticLongitudeZero}{\ensuremath{2.0}}
\newcommand{\RunSixInjBHBOneInclinationZero}{\ensuremath{0.6435011087932843}}
\newcommand{\RunSixInjBHBOneInitialOrbitalPhaseZero}{\ensuremath{1.0}}
\newcommand{\RunSixInjBHBOneLuminosityDistanceZero}{\ensuremath{47.6}}
\newcommand{\RunSixInjBHBOneMergerTimeOrInitialOrbitalFrequencyZero}{\ensuremath{30.0}}
\newcommand{\RunSixInjBHBOnePolarizationZero}{\ensuremath{1.7}}
\newcommand{\RunSixInjBHBOneRedshiftedMassOneZero}{\ensuremath{4.5}}
\newcommand{\RunSixInjBHBOneRedshiftedMassTwoZero}{\ensuremath{1.5}}
\newcommand{\RunSixInjLPFOneDeltavZero}{\ensuremath{3e-12}}
\newcommand{\RunSixInjLPFOneInitialTimeZero}{\ensuremath{108785.626}}
\newcommand{\RunSixInjLPFOneTauFallZero}{\ensuremath{1800.0}}
\newcommand{\RunSixInjLPFOneTauRiseZero}{\ensuremath{1500.0}}
%File created on 29-05-2023 17:19:55

%This file is machine generated. Do *not* edit it manually if you want reproducibility.
%Hashed ..//run7/config_inj.yaml: fe5cf340c8b3ac2b124df4a3c03ccf54
%Hashed ..//run7/config_pe.yaml: d831b1f7a74c2df07530ac47d6db6a99
%Hashed ..//run7/physical_posterior.dat: 49f7cb7d51a6c257d17b7a0bc2596cf5
\newcommand{\RunSevenBHBOneDimensionlessSpinOneZero}{\ensuremath{0.26^{+0.16}_{-0.04}}}
\newcommand{\RunSevenBHBOneDimensionlessSpinTwoZero}{\ensuremath{0.95^{+0.05}_{-1.77}}}
\newcommand{\RunSevenBHBOneEclipticLatitudeZero}{\ensuremath{0.3^{+0.4}_{-0.1}}}
\newcommand{\RunSevenBHBOneEclipticLongitudeZero}{\ensuremath{1.95^{+0.29}_{-0.07}}}
\newcommand{\RunSevenBHBOneInclinationZero}{\ensuremath{0.8^{+0.3}_{-0.6}}}
\newcommand{\RunSevenBHBOneInitialOrbitalPhaseZero}{\ensuremath{1.5^{+1.4}_{-1.3}}}
\newcommand{\RunSevenBHBOneLuminosityDistanceZero}{\ensuremath{36^{+12}_{-12}}}
\newcommand{\RunSevenBHBOneMergerTimeOrInitialOrbitalFrequencyZero}{\ensuremath{29.85^{+0.06}_{-0.06}}}
\newcommand{\RunSevenBHBOnePolarizationZero}{\ensuremath{1.6^{+1.3}_{-1.4}}}
\newcommand{\RunSevenBHBOneRedshiftedMassOneZero}{\ensuremath{4.4^{+0.3}_{-0.1}}}
\newcommand{\RunSevenBHBOneRedshiftedMassTwoZero}{\ensuremath{1.4^{+0.2}_{-0.5}}}
\newcommand{\RunSevenInjBHBOneDimensionlessSpinOneZero}{\ensuremath{0.4}}
\newcommand{\RunSevenInjBHBOneDimensionlessSpinTwoZero}{\ensuremath{0.3}}
\newcommand{\RunSevenInjBHBOneEclipticLatitudeZero}{\ensuremath{0.30}}
\newcommand{\RunSevenInjBHBOneEclipticLongitudeZero}{\ensuremath{2.0}}
\newcommand{\RunSevenInjBHBOneInclinationZero}{\ensuremath{0.64}}
\newcommand{\RunSevenInjBHBOneInitialOrbitalPhaseZero}{\ensuremath{1.0}}
\newcommand{\RunSevenInjBHBOneLuminosityDistanceZero}{\ensuremath{47.6}}
\newcommand{\RunSevenInjBHBOneMergerTimeOrInitialOrbitalFrequencyZero}{\ensuremath{30.0}}
\newcommand{\RunSevenInjBHBOnePolarizationZero}{\ensuremath{1.7}}
\newcommand{\RunSevenInjBHBOneRedshiftedMassOneZero}{\ensuremath{4.5}}
\newcommand{\RunSevenInjBHBOneRedshiftedMassTwoZero}{\ensuremath{1.5}}
\newcommand{\RunSevenInjSLOneAmplitudeDisplacementZero}{\ensuremath{0.0948}}
\newcommand{\RunSevenInjSLOneBetaZero}{\ensuremath{40.0}}
\newcommand{\RunSevenInjSLOneInitialTimeZero}{\ensuremath{29.48}}
%File created on 29-05-2023 17:52:11

%This file is machine generated. Do *not* edit it manually if you want reproducibility.
%Hashed ..//run8/config_inj.yaml: d3d60f33422381c6ef0adda214a68d29
%Hashed ..//run8/config_pe.yaml: bb2da91761ebf69dbeadd2fd4c9fcfe1
%Hashed ..//run8/physical_posterior.dat: 91aeaeda9e9ee9cf17486bf1e57e5316
\newcommand{\RunEightBHBOneDimensionlessSpinOneZero}{\ensuremath{0.4^{+0.1}_{-0.1}}}
\newcommand{\RunEightBHBOneDimensionlessSpinTwoZero}{\ensuremath{0.3^{+0.6}_{-1.0}}}
\newcommand{\RunEightBHBOneEclipticLatitudeZero}{\ensuremath{0.3^{+0.5}_{-0.1}}}
\newcommand{\RunEightBHBOneEclipticLongitudeZero}{\ensuremath{1.99^{+0.25}_{-0.10}}}
\newcommand{\RunEightBHBOneInclinationZero}{\ensuremath{0.8^{+0.3}_{-0.6}}}
\newcommand{\RunEightBHBOneInitialOrbitalPhaseZero}{\ensuremath{1.5^{+1.5}_{-1.4}}}
\newcommand{\RunEightBHBOneLuminosityDistanceZero}{\ensuremath{44^{+15}_{-15}}}
\newcommand{\RunEightBHBOneMergerTimeOrInitialOrbitalFrequencyZero}{\ensuremath{30.01^{+0.09}_{-0.08}}}
\newcommand{\RunEightBHBOnePolarizationZero}{\ensuremath{1.6^{+1.3}_{-1.4}}}
\newcommand{\RunEightBHBOneRedshiftedMassOneZero}{\ensuremath{4.5^{+0.2}_{-0.2}}}
\newcommand{\RunEightBHBOneRedshiftedMassTwoZero}{\ensuremath{1.5^{+0.3}_{-0.3}}}
\newcommand{\RunEightInjBHBOneDimensionlessSpinOneZero}{\ensuremath{0.4}}
\newcommand{\RunEightInjBHBOneDimensionlessSpinTwoZero}{\ensuremath{0.3}}
\newcommand{\RunEightInjBHBOneEclipticLatitudeZero}{\ensuremath{0.3}}
\newcommand{\RunEightInjBHBOneEclipticLongitudeZero}{\ensuremath{2.0}}
\newcommand{\RunEightInjBHBOneInclinationZero}{\ensuremath{0.6}}
\newcommand{\RunEightInjBHBOneInitialOrbitalPhaseZero}{\ensuremath{1.0}}
\newcommand{\RunEightInjBHBOneLuminosityDistanceZero}{\ensuremath{47.6}}
\newcommand{\RunEightInjBHBOneMergerTimeOrInitialOrbitalFrequencyZero}{\ensuremath{30.0}}
\newcommand{\RunEightInjBHBOnePolarizationZero}{\ensuremath{1.7}}
\newcommand{\RunEightInjBHBOneRedshiftedMassOneZero}{\ensuremath{4.5}}
\newcommand{\RunEightInjBHBOneRedshiftedMassTwoZero}{\ensuremath{1.5}}
\newcommand{\RunEightInjSLOneAmplitudeDisplacementZero}{\ensuremath{200.0}}
\newcommand{\RunEightInjSLOneBetaZero}{\ensuremath{5.0}}
\newcommand{\RunEightInjSLOneInitialTimeZero}{\ensuremath{29.507}}
%File created on 29-05-2023 17:58:05

%This file is machine generated. Do *not* edit it manually if you want reproducibility.
%Hashed ..//run9/config_inj.yaml: aacec94b12e58101646c4839aeffac43
%Hashed ..//run9/config_pe.yaml: aacec94b12e58101646c4839aeffac43
%Hashed ..//run9/physical_posterior.dat: 037c96b36da25279ed56abf75d7659ec
\newcommand{\RunNineBHBOneDimensionlessSpinOneZero}{\ensuremath{0.4^{+0.1}_{-0.1}}}
\newcommand{\RunNineBHBOneDimensionlessSpinTwoZero}{\ensuremath{0.3^{+0.6}_{-1.0}}}
\newcommand{\RunNineBHBOneEclipticLatitudeZero}{\ensuremath{0.3^{+0.6}_{-0.1}}}
\newcommand{\RunNineBHBOneEclipticLongitudeZero}{\ensuremath{2.0^{+0.3}_{-0.1}}}
\newcommand{\RunNineBHBOneInclinationZero}{\ensuremath{0.8^{+0.3}_{-0.6}}}
\newcommand{\RunNineBHBOneInitialOrbitalPhaseZero}{\ensuremath{1.6^{+1.4}_{-1.3}}}
\newcommand{\RunNineBHBOneLuminosityDistanceZero}{\ensuremath{44^{+15}_{-15}}}
\newcommand{\RunNineBHBOneMergerTimeOrInitialOrbitalFrequencyZero}{\ensuremath{30.00^{+0.10}_{-0.08}}}
\newcommand{\RunNineBHBOnePolarizationZero}{\ensuremath{1.6^{+1.3}_{-1.4}}}
\newcommand{\RunNineBHBOneRedshiftedMassOneZero}{\ensuremath{4.5^{+0.2}_{-0.2}}}
\newcommand{\RunNineBHBOneRedshiftedMassTwoZero}{\ensuremath{1.5^{+0.3}_{-0.3}}}
\newcommand{\RunNineLPFOneDeltavZero}{\ensuremath{3.0^{+0.2}_{-0.1}}}
\newcommand{\RunNineLPFOneInitialTimeZero}{\ensuremath{30.219^{+0.006}_{-0.006}}}
\newcommand{\RunNineLPFOneTauFallZero}{\ensuremath{1647^{+584}_{-426}}}
\newcommand{\RunNineLPFOneTauRiseZero}{\ensuremath{1637^{+594}_{-417}}}
\newcommand{\RunNineInjBHBOneDimensionlessSpinOneZero}{\ensuremath{0.4}}
\newcommand{\RunNineInjBHBOneDimensionlessSpinTwoZero}{\ensuremath{0.3}}
\newcommand{\RunNineInjBHBOneEclipticLatitudeZero}{\ensuremath{0.30}}
\newcommand{\RunNineInjBHBOneEclipticLongitudeZero}{\ensuremath{2.0}}
\newcommand{\RunNineInjBHBOneInclinationZero}{\ensuremath{0.6}}
\newcommand{\RunNineInjBHBOneInitialOrbitalPhaseZero}{\ensuremath{1.0}}
\newcommand{\RunNineInjBHBOneLuminosityDistanceZero}{\ensuremath{47.6}}
\newcommand{\RunNineInjBHBOneMergerTimeOrInitialOrbitalFrequencyZero}{\ensuremath{30.0}}
\newcommand{\RunNineInjBHBOnePolarizationZero}{\ensuremath{1.7}}
\newcommand{\RunNineInjBHBOneRedshiftedMassOneZero}{\ensuremath{4.5}}
\newcommand{\RunNineInjBHBOneRedshiftedMassTwoZero}{\ensuremath{1.5}}
\newcommand{\RunNineInjLPFOneDeltavZero}{\ensuremath{3.0}}
\newcommand{\RunNineInjLPFOneInitialTimeZero}{\ensuremath{30.21}}
\newcommand{\RunNineInjLPFOneTauFallZero}{\ensuremath{1800.0}}
\newcommand{\RunNineInjLPFOneTauRiseZero}{\ensuremath{1500.0}}
%File created on 29-05-2023 17:16:02

%This file is machine generated. Do *not* edit it manually if you want reproducibility.
%Hashed ..//run10/config_inj.yaml: fe5cf340c8b3ac2b124df4a3c03ccf54
%Hashed ..//run10/config_pe.yaml: fe5cf340c8b3ac2b124df4a3c03ccf54
%Hashed ..//run10/physical_posterior.dat: 1669f046535cc3fb52b750c9f1d15f51
\newcommand{\RunTenBHBOneDimensionlessSpinOneZero}{\ensuremath{0.4^{+0.1}_{-0.1}}}
\newcommand{\RunTenBHBOneDimensionlessSpinTwoZero}{\ensuremath{0.3^{+0.6}_{-1.0}}}
\newcommand{\RunTenBHBOneEclipticLatitudeZero}{\ensuremath{0.3^{+0.6}_{-0.1}}}
\newcommand{\RunTenBHBOneEclipticLongitudeZero}{\ensuremath{2.0^{+0.2}_{-0.1}}}
\newcommand{\RunTenBHBOneInclinationZero}{\ensuremath{0.8^{+0.3}_{-0.6}}}
\newcommand{\RunTenBHBOneInitialOrbitalPhaseZero}{\ensuremath{1.6^{+1.4}_{-1.3}}}
\newcommand{\RunTenBHBOneLuminosityDistanceZero}{\ensuremath{44^{+15}_{-15}}}
\newcommand{\RunTenBHBOneMergerTimeOrInitialOrbitalFrequencyZero}{\ensuremath{30.01^{+0.10}_{-0.08}}}
\newcommand{\RunTenBHBOnePolarizationZero}{\ensuremath{1.6^{+1.3}_{-1.4}}}
\newcommand{\RunTenBHBOneRedshiftedMassOneZero}{\ensuremath{4.5^{+0.2}_{-0.2}}}
\newcommand{\RunTenBHBOneRedshiftedMassTwoZero}{\ensuremath{1.5^{+0.3}_{-0.3}}}
\newcommand{\RunTenSLOneAmplitudeDisplacementZero}{\ensuremath{0.095^{+0.007}_{-0.007}}}
\newcommand{\RunTenSLOneBetaZero}{\ensuremath{41^{+22}_{-23}}}
\newcommand{\RunTenSLOneInitialTimeZero}{\ensuremath{29.48^{+0.01}_{-0.01}}}
\newcommand{\RunTenInjBHBOneDimensionlessSpinOneZero}{\ensuremath{0.4}}
\newcommand{\RunTenInjBHBOneDimensionlessSpinTwoZero}{\ensuremath{0.3}}
\newcommand{\RunTenInjBHBOneEclipticLatitudeZero}{\ensuremath{0.3}}
\newcommand{\RunTenInjBHBOneEclipticLongitudeZero}{\ensuremath{2.0}}
\newcommand{\RunTenInjBHBOneInclinationZero}{\ensuremath{0.6}}
\newcommand{\RunTenInjBHBOneInitialOrbitalPhaseZero}{\ensuremath{1.0}}
\newcommand{\RunTenInjBHBOneLuminosityDistanceZero}{\ensuremath{47.6}}
\newcommand{\RunTenInjBHBOneMergerTimeOrInitialOrbitalFrequencyZero}{\ensuremath{30.0}}
\newcommand{\RunTenInjBHBOnePolarizationZero}{\ensuremath{1.7}}
\newcommand{\RunTenInjBHBOneRedshiftedMassOneZero}{\ensuremath{4.5}}
\newcommand{\RunTenInjBHBOneRedshiftedMassTwoZero}{\ensuremath{1.5}}
\newcommand{\RunTenInjSLOneAmplitudeDisplacementZero}{\ensuremath{0.0948}}
\newcommand{\RunTenInjSLOneBetaZero}{\ensuremath{40.0}}
\newcommand{\RunTenInjSLOneInitialTimeZero}{\ensuremath{29.48}}
%File created on 29-05-2023 17:39:53

%This file is machine generated. Do *not* edit it manually if you want reproducibility.
%Hashed ..//run11/config_inj.yaml: d3d60f33422381c6ef0adda214a68d29
%Hashed ..//run11/config_pe.yaml: d3d60f33422381c6ef0adda214a68d29
%Hashed ..//run11/physical_posterior.dat: aee416012a8a54aacae1d76bfaf2fbf6
\newcommand{\RunElevenBHBOneDimensionlessSpinOneZero}{\ensuremath{0.4^{+0.1}_{-0.1}}}
\newcommand{\RunElevenBHBOneDimensionlessSpinTwoZero}{\ensuremath{0.3^{+0.6}_{-1.0}}}
\newcommand{\RunElevenBHBOneEclipticLatitudeZero}{\ensuremath{0.3^{+0.5}_{-0.1}}}
\newcommand{\RunElevenBHBOneEclipticLongitudeZero}{\ensuremath{1.99^{+0.25}_{-0.09}}}
\newcommand{\RunElevenBHBOneInclinationZero}{\ensuremath{0.8^{+0.3}_{-0.6}}}
\newcommand{\RunElevenBHBOneInitialOrbitalPhaseZero}{\ensuremath{1.6^{+1.4}_{-1.3}}}
\newcommand{\RunElevenBHBOneLuminosityDistanceZero}{\ensuremath{44^{+15}_{-14}}}
\newcommand{\RunElevenBHBOneMergerTimeOrInitialOrbitalFrequencyZero}{\ensuremath{30.01^{+0.09}_{-0.08}}}
\newcommand{\RunElevenBHBOnePolarizationZero}{\ensuremath{1.6^{+1.3}_{-1.4}}}
\newcommand{\RunElevenBHBOneRedshiftedMassOneZero}{\ensuremath{4.5^{+0.2}_{-0.2}}}
\newcommand{\RunElevenBHBOneRedshiftedMassTwoZero}{\ensuremath{1.5^{+0.3}_{-0.3}}}
\newcommand{\RunElevenSLOneAmplitudeDisplacementZero}{\ensuremath{216^{+83}_{-73}}}
\newcommand{\RunElevenSLOneBetaZero}{\ensuremath{16^{+13}_{-14}}}
\newcommand{\RunElevenSLOneInitialTimeZero}{\ensuremath{29.502^{+0.009}_{-0.008}}}
\newcommand{\RunElevenInjBHBOneDimensionlessSpinOneZero}{\ensuremath{0.4}}
\newcommand{\RunElevenInjBHBOneDimensionlessSpinTwoZero}{\ensuremath{0.3}}
\newcommand{\RunElevenInjBHBOneEclipticLatitudeZero}{\ensuremath{0.3}}
\newcommand{\RunElevenInjBHBOneEclipticLongitudeZero}{\ensuremath{2.0}}
\newcommand{\RunElevenInjBHBOneInclinationZero}{\ensuremath{0.6}}
\newcommand{\RunElevenInjBHBOneInitialOrbitalPhaseZero}{\ensuremath{1.0}}
\newcommand{\RunElevenInjBHBOneLuminosityDistanceZero}{\ensuremath{47.6}}
\newcommand{\RunElevenInjBHBOneMergerTimeOrInitialOrbitalFrequencyZero}{\ensuremath{30.0}}
\newcommand{\RunElevenInjBHBOnePolarizationZero}{\ensuremath{1.7}}
\newcommand{\RunElevenInjBHBOneRedshiftedMassOneZero}{\ensuremath{4.5}}
\newcommand{\RunElevenInjBHBOneRedshiftedMassTwoZero}{\ensuremath{1.5}}
\newcommand{\RunElevenInjSLOneAmplitudeDisplacementZero}{\ensuremath{200.0}}
\newcommand{\RunElevenInjSLOneBetaZero}{\ensuremath{5.0}}
\newcommand{\RunElevenInjSLOneInitialTimeZero}{\ensuremath{29.507}}
%File created on 29-05-2023 17:55:37

%This file is machine generated. Do *not* edit it manually if you want reproducibility.
%Hashed ../results/final_runs//run12/config_inj.yaml: 6aa68947c0b25dc43079c1ee25399887
%Hashed ../results/final_runs//run12/config_pe.yaml: cf892e9b59b303af6f0eb57eafd13105
%Hashed ../results/final_runs//run12/physical_posterior.dat: 10adb55234383d49791193094af477f2
\newcommand{\RunTwelveSLOneAmplitudeDisplacementZero}{\ensuremath{22.42^{+20.45}_{-6.98}}}
\newcommand{\RunTwelveSLOneBetaZero}{\ensuremath{20.42^{+8.76}_{-16.56}}}
\newcommand{\RunTwelveSLOneInitialTimeZero}{\ensuremath{3593^{+217}_{-214}}}
\newcommand{\RunTwelveBHBOneDimensionlessSpinOneZero}{\ensuremath{0.40^{+0.04}_{-0.04}}}
\newcommand{\RunTwelveBHBOneDimensionlessSpinTwoZero}{\ensuremath{0.3^{+0.1}_{-0.1}}}
\newcommand{\RunTwelveBHBOneEclipticLatitudeZero}{\ensuremath{0.3^{+0.4}_{-0.2}}}
\newcommand{\RunTwelveBHBOneEclipticLongitudeZero}{\ensuremath{1.95^{+0.21}_{-0.08}}}
\newcommand{\RunTwelveBHBOneInclinationZero}{\ensuremath{0.8^{+0.3}_{-0.5}}}
\newcommand{\RunTwelveBHBOneInitialOrbitalPhaseZero}{\ensuremath{1.2^{+0.6}_{-0.7}}}
\newcommand{\RunTwelveBHBOneLuminosityDistanceZero}{\ensuremath{16.0^{+4.6}_{-5.3}}}
\newcommand{\RunTwelveBHBOneMergerTimeOrInitialOrbitalFrequencyZero}{\ensuremath{3621^{+43}_{-49}}}
\newcommand{\RunTwelveBHBOnePolarizationZero}{\ensuremath{3.5^{+0.7}_{-0.7}}}
\newcommand{\RunTwelveBHBOneRedshiftedMassOneZero}{\ensuremath{2.00^{+0.03}_{-0.03}}}
\newcommand{\RunTwelveBHBOneRedshiftedMassTwoZero}{\ensuremath{1.10^{+0.03}_{-0.03}}}
\newcommand{\RunTwelveInjBHBOneDimensionlessSpinOneZero}{\ensuremath{0.4}}
\newcommand{\RunTwelveInjBHBOneDimensionlessSpinTwoZero}{\ensuremath{0.3}}
\newcommand{\RunTwelveInjBHBOneEclipticLatitudeZero}{\ensuremath{0.30469265401539747}}
\newcommand{\RunTwelveInjBHBOneEclipticLongitudeZero}{\ensuremath{2.0}}
\newcommand{\RunTwelveInjBHBOneInclinationZero}{\ensuremath{0.8}}
\newcommand{\RunTwelveInjBHBOneInitialOrbitalPhaseZero}{\ensuremath{1.0}}
\newcommand{\RunTwelveInjBHBOneLuminosityDistanceZero}{\ensuremath{16.0}}
\newcommand{\RunTwelveInjBHBOneMergerTimeOrInitialOrbitalFrequencyZero}{\ensuremath{3600.0}}
\newcommand{\RunTwelveInjBHBOnePolarizationZero}{\ensuremath{3.3}}
\newcommand{\RunTwelveInjBHBOneRedshiftedMassOneZero}{\ensuremath{2.0}}
\newcommand{\RunTwelveInjBHBOneRedshiftedMassTwoZero}{\ensuremath{1.1}}
%File created on 07-05-2023 01:35:50
%This file is machine generated. Do *not* edit it manually if you want reproducibility.
%Hashed ../results/final_runs//run13/config_inj.yaml: 29d3ad6ecdc370015f2fac7921fd02ea
%Hashed ../results/final_runs//run13/config_pe.yaml: 10c6ab4eb48803899c71a896d7ab9353
%Hashed ../results/final_runs//run13/physical_posterior.dat: 18bf5ca20e148010212139c9f940f53c
\newcommand{\RunThirteenSLOneAmplitudeDisplacementZero}{\ensuremath{0.0098^{+0.0011}_{-0.0003}}}
\newcommand{\RunThirteenSLOneBetaZero}{\ensuremath{81^{+35}_{-64}}}
\newcommand{\RunThirteenSLOneInitialTimeZero}{\ensuremath{3531^{+238}_{-191}}}
\newcommand{\RunThirteenBHBOneDimensionlessSpinOneZero}{\ensuremath{0.39^{+0.04}_{-0.05}}}
\newcommand{\RunThirteenBHBOneDimensionlessSpinTwoZero}{\ensuremath{0.3^{+0.1}_{-0.1}}}
\newcommand{\RunThirteenBHBOneEclipticLatitudeZero}{\ensuremath{0.3^{+0.7}_{-0.1}}}
\newcommand{\RunThirteenBHBOneEclipticLongitudeZero}{\ensuremath{1.96^{+0.73}_{-0.07}}}
\newcommand{\RunThirteenBHBOneInclinationZero}{\ensuremath{0.8^{+0.3}_{-0.5}}}
\newcommand{\RunThirteenBHBOneInitialOrbitalPhaseZero}{\ensuremath{1.2^{+0.6}_{-0.7}}}
\newcommand{\RunThirteenBHBOneLuminosityDistanceZero}{\ensuremath{16.3^{+4.5}_{-4.3}}}
\newcommand{\RunThirteenBHBOneMergerTimeOrInitialOrbitalFrequencyZero}{\ensuremath{3619^{+43}_{-47}}}
\newcommand{\RunThirteenBHBOnePolarizationZero}{\ensuremath{3.5^{+0.6}_{-0.7}}}
\newcommand{\RunThirteenBHBOneRedshiftedMassOneZero}{\ensuremath{1.99^{+0.03}_{-0.03}}}
\newcommand{\RunThirteenBHBOneRedshiftedMassTwoZero}{\ensuremath{1.11^{+0.03}_{-0.03}}}
\newcommand{\RunThirteenInjBHBOneDimensionlessSpinOneZero}{\ensuremath{0.4}}
\newcommand{\RunThirteenInjBHBOneDimensionlessSpinTwoZero}{\ensuremath{0.3}}
\newcommand{\RunThirteenInjBHBOneEclipticLatitudeZero}{\ensuremath{0.30469265401539747}}
\newcommand{\RunThirteenInjBHBOneEclipticLongitudeZero}{\ensuremath{2.0}}
\newcommand{\RunThirteenInjBHBOneInclinationZero}{\ensuremath{0.8}}
\newcommand{\RunThirteenInjBHBOneInitialOrbitalPhaseZero}{\ensuremath{1.0}}
\newcommand{\RunThirteenInjBHBOneLuminosityDistanceZero}{\ensuremath{16.0}}
\newcommand{\RunThirteenInjBHBOneMergerTimeOrInitialOrbitalFrequencyZero}{\ensuremath{3600.0}}
\newcommand{\RunThirteenInjBHBOnePolarizationZero}{\ensuremath{3.3}}
\newcommand{\RunThirteenInjBHBOneRedshiftedMassOneZero}{\ensuremath{2.0}}
\newcommand{\RunThirteenInjBHBOneRedshiftedMassTwoZero}{\ensuremath{1.1}}
%File created on 07-05-2023 01:35:57
%This file is machine generated. Do *not* edit it manually if you want reproducibility.
%Hashed ../results/final_runs//run14/config_inj.yaml: 8c9a136a3ee5e450941e4955eee2a6f9
%Hashed ../results/final_runs//run14/config_pe.yaml: 89365c51040d86eb307cf315761e6321
%Hashed ../results/final_runs//run14/physical_posterior.dat: 2630a1e23ba03c8353a498cf46ea625f
\newcommand{\RunFourteenLPFOneDeltavZero}{\ensuremath{0.301^{+0.004}_{-0.001}}}
\newcommand{\RunFourteenLPFOneInitialTimeZero}{\ensuremath{1773^{+233}_{-192}}}
\newcommand{\RunFourteenLPFOneTauFallZero}{\ensuremath{2089.69^{+9.44}_{-33.23}}}
\newcommand{\RunFourteenLPFOneTauRiseZero}{\ensuremath{1889.88^{+9.30}_{-31.71}}}
\newcommand{\RunFourteenBHBOneDimensionlessSpinOneZero}{\ensuremath{0.43^{+0.04}_{-0.04}}}
\newcommand{\RunFourteenBHBOneDimensionlessSpinTwoZero}{\ensuremath{0.2^{+0.1}_{-0.1}}}
\newcommand{\RunFourteenBHBOneEclipticLatitudeZero}{\ensuremath{0.3^{+0.3}_{-0.2}}}
\newcommand{\RunFourteenBHBOneEclipticLongitudeZero}{\ensuremath{1.96^{+0.20}_{-0.06}}}
\newcommand{\RunFourteenBHBOneInclinationZero}{\ensuremath{0.8^{+0.3}_{-0.5}}}
\newcommand{\RunFourteenBHBOneInitialOrbitalPhaseZero}{\ensuremath{1.2^{+0.6}_{-0.7}}}
\newcommand{\RunFourteenBHBOneLuminosityDistanceZero}{\ensuremath{15.9^{+4.5}_{-5.4}}}
\newcommand{\RunFourteenBHBOneMergerTimeOrInitialOrbitalFrequencyZero}{\ensuremath{3629^{+38}_{-55}}}
\newcommand{\RunFourteenBHBOnePolarizationZero}{\ensuremath{3.4^{+0.7}_{-0.6}}}
\newcommand{\RunFourteenBHBOneRedshiftedMassOneZero}{\ensuremath{2.02^{+0.03}_{-0.03}}}
\newcommand{\RunFourteenBHBOneRedshiftedMassTwoZero}{\ensuremath{1.08^{+0.03}_{-0.03}}}
\newcommand{\RunFourteenInjBHBOneDimensionlessSpinOneZero}{\ensuremath{0.4}}
\newcommand{\RunFourteenInjBHBOneDimensionlessSpinTwoZero}{\ensuremath{0.3}}
\newcommand{\RunFourteenInjBHBOneEclipticLatitudeZero}{\ensuremath{0.30469265401539747}}
\newcommand{\RunFourteenInjBHBOneEclipticLongitudeZero}{\ensuremath{2.0}}
\newcommand{\RunFourteenInjBHBOneInclinationZero}{\ensuremath{0.8}}
\newcommand{\RunFourteenInjBHBOneInitialOrbitalPhaseZero}{\ensuremath{1.0}}
\newcommand{\RunFourteenInjBHBOneLuminosityDistanceZero}{\ensuremath{16.0}}
\newcommand{\RunFourteenInjBHBOneMergerTimeOrInitialOrbitalFrequencyZero}{\ensuremath{3600.0}}
\newcommand{\RunFourteenInjBHBOnePolarizationZero}{\ensuremath{3.3}}
\newcommand{\RunFourteenInjBHBOneRedshiftedMassOneZero}{\ensuremath{2.0}}
\newcommand{\RunFourteenInjBHBOneRedshiftedMassTwoZero}{\ensuremath{1.1}}
%File created on 07-05-2023 01:36:07

%This file is machine generated. Do *not* edit it manually if you want reproducibility.
%Hashed ..//run15/config_inj.yaml: affeacc69e16ae90d898556388cf95c1
%Hashed ..//run15/config_pe.yaml: 39abfd624bfdaea2733b9439e8c75b1e
%Hashed ..//run15/physical_posterior.dat: 36d66a2e8fd6b0f637a9b714128ec269
\newcommand{\RunFifteenSLOneAmplitudeDisplacementZero}{\ensuremath{672^{+536}_{-278}}}
\newcommand{\RunFifteenSLOneAmplitudeDisplacementOne}{\ensuremath{1336^{+690}_{-578}}}
\newcommand{\RunFifteenSLOneBetaZero}{\ensuremath{69^{+60}_{-29}}}
\newcommand{\RunFifteenSLOneBetaOne}{\ensuremath{77^{+16}_{-15}}}
\newcommand{\RunFifteenSLOneInitialTimeZero}{\ensuremath{11.997^{+0.005}_{-0.007}}}
\newcommand{\RunFifteenSLOneInitialTimeOne}{\ensuremath{12.015^{+0.010}_{-0.012}}}
\newcommand{\RunFifteenInjSLOneAmplitudeDisplacementZero}{\ensuremath{542.0}}
\newcommand{\RunFifteenInjSLOneAmplitudeDisplacementOne}{\ensuremath{1420.0}}
\newcommand{\RunFifteenInjSLOneBetaZero}{\ensuremath{40.0}}
\newcommand{\RunFifteenInjSLOneBetaOne}{\ensuremath{80.0}}
\newcommand{\RunFifteenInjSLOneInitialTimeZero}{\ensuremath{12.0}}
\newcommand{\RunFifteenInjSLOneInitialTimeOne}{\ensuremath{12.0}}
%File created on 29-05-2023 18:13:01

%This file is machine generated. Do *not* edit it manually if you want reproducibility.
%Hashed ../results/final_runs//run16/config_inj.yaml: 856b103d6343b771344209e7ed8032ed
%Hashed ../results/final_runs//run16/config_pe.yaml: cb9589dc6fdf2a3ec033e4c707edcfc6
%Hashed ../results/final_runs//run16/physical_posterior.dat: 1a51f7e181546fa760261b64b3e9a929
\newcommand{\RunSixteenSLOneAmplitudeDisplacementZero}{\ensuremath{1501^{+191}_{-153}}}
\newcommand{\RunSixteenSLOneAmplitudeDisplacementOne}{\ensuremath{1043^{+69}_{-69}}}
\newcommand{\RunSixteenSLOneBetaZero}{\ensuremath{56.0^{+3.8}_{-3.7}}}
\newcommand{\RunSixteenSLOneBetaOne}{\ensuremath{104.7^{+3.4}_{-3.2}}}
\newcommand{\RunSixteenSLOneInitialTimeZero}{\ensuremath{43206.1^{+2.8}_{-2.9}}}
\newcommand{\RunSixteenSLOneInitialTimeOne}{\ensuremath{43144^{+12}_{-15}}}
\newcommand{\RunSixteenInjSLOneAmplitudeDisplacementZero}{\ensuremath{542.0}}
\newcommand{\RunSixteenInjSLOneAmplitudeDisplacementOne}{\ensuremath{1420.0}}
\newcommand{\RunSixteenInjSLOneBetaZero}{\ensuremath{40.0}}
\newcommand{\RunSixteenInjSLOneBetaOne}{\ensuremath{80.0}}
\newcommand{\RunSixteenInjSLOneInitialTimeZero}{\ensuremath{43200.0}}
\newcommand{\RunSixteenInjSLOneInitialTimeOne}{\ensuremath{43200.0}}
%File created on 07-05-2023 01:36:22
%This file is machine generated. Do *not* edit it manually if you want reproducibility.
%Hashed ../results/final_runs//run17/config_inj.yaml: 856b103d6343b771344209e7ed8032ed
%Hashed ../results/final_runs//run17/config_pe.yaml: b1bc1cd9df542d069374dc554f6d1867
%Hashed ../results/final_runs//run17/physical_posterior.dat: 38dbb2485d1be021b19a43d1189a76a1
\newcommand{\RunSeventeenSLOneAmplitudeDisplacementZero}{\ensuremath{2602^{+183}_{-164}}}
\newcommand{\RunSeventeenSLOneAmplitudeDisplacementOne}{\ensuremath{391^{+110}_{-101}}}
\newcommand{\RunSeventeenSLOneBetaZero}{\ensuremath{106.7^{+7.6}_{-7.1}}}
\newcommand{\RunSeventeenSLOneBetaOne}{\ensuremath{54.0^{+4.0}_{-4.7}}}
\newcommand{\RunSeventeenSLOneInitialTimeZero}{\ensuremath{43191.9^{+5.3}_{-5.3}}}
\newcommand{\RunSeventeenSLOneInitialTimeOne}{\ensuremath{43192.4^{+8.6}_{-8.0}}}
\newcommand{\RunSeventeenInjSLOneAmplitudeDisplacementZero}{\ensuremath{542.0}}
\newcommand{\RunSeventeenInjSLOneAmplitudeDisplacementOne}{\ensuremath{1420.0}}
\newcommand{\RunSeventeenInjSLOneBetaZero}{\ensuremath{40.0}}
\newcommand{\RunSeventeenInjSLOneBetaOne}{\ensuremath{80.0}}
\newcommand{\RunSeventeenInjSLOneInitialTimeZero}{\ensuremath{43200.0}}
\newcommand{\RunSeventeenInjSLOneInitialTimeOne}{\ensuremath{43200.0}}
%File created on 07-05-2023 01:36:29
%This file is machine generated. Do *not* edit it manually if you want reproducibility.
%Hashed ../results/final_runs//run18/config_inj.yaml: 856b103d6343b771344209e7ed8032ed
%Hashed ../results/final_runs//run18/config_pe.yaml: e7fdf14eef4a80a66cf6e94351cf187f
%Hashed ../results/final_runs//run18/physical_posterior.dat: 8d4488ea4811cd091a6762ee5be74aa2
\newcommand{\RunEighteenSLOneAmplitudeDisplacementZero}{\ensuremath{2206^{+84}_{-86}}}
\newcommand{\RunEighteenSLOneBetaZero}{\ensuremath{58.1^{+2.0}_{-2.0}}}
\newcommand{\RunEighteenSLOneInitialTimeZero}{\ensuremath{43194.9^{+2.1}_{-2.1}}}
\newcommand{\RunEighteenInjSLOneAmplitudeDisplacementZero}{\ensuremath{542.0}}
\newcommand{\RunEighteenInjSLOneAmplitudeDisplacementOne}{\ensuremath{1420.0}}
\newcommand{\RunEighteenInjSLOneBetaZero}{\ensuremath{40.0}}
\newcommand{\RunEighteenInjSLOneBetaOne}{\ensuremath{80.0}}
\newcommand{\RunEighteenInjSLOneInitialTimeZero}{\ensuremath{43200.0}}
\newcommand{\RunEighteenInjSLOneInitialTimeOne}{\ensuremath{43200.0}}
%File created on 07-05-2023 01:36:36
%This file is machine generated. Do *not* edit it manually if you want reproducibility.
%Hashed ../results/final_runs//run19/config_inj.yaml: 856b103d6343b771344209e7ed8032ed
%Hashed ../results/final_runs//run19/config_pe.yaml: 124b46df10eca7550cec6d9d8658afed
%Hashed ../results/final_runs//run19/physical_posterior.dat: d1ed702a71e2304a2ddb07c410f7d895
\newcommand{\RunNineteenSLOneAmplitudeDisplacementZero}{\ensuremath{1861^{+58}_{-58}}}
\newcommand{\RunNineteenSLOneBetaZero}{\ensuremath{80.8^{+1.5}_{-1.5}}}
\newcommand{\RunNineteenSLOneInitialTimeZero}{\ensuremath{43191.8^{+2.2}_{-2.2}}}
\newcommand{\RunNineteenInjSLOneAmplitudeDisplacementZero}{\ensuremath{542.0}}
\newcommand{\RunNineteenInjSLOneAmplitudeDisplacementOne}{\ensuremath{1420.0}}
\newcommand{\RunNineteenInjSLOneBetaZero}{\ensuremath{40.0}}
\newcommand{\RunNineteenInjSLOneBetaOne}{\ensuremath{80.0}}
\newcommand{\RunNineteenInjSLOneInitialTimeZero}{\ensuremath{43200.0}}
\newcommand{\RunNineteenInjSLOneInitialTimeOne}{\ensuremath{43200.0}}

%File created on 07-05-2023 01:32:00
%This file is machine generated. Do *not* edit it manually if you want reproducibility.
%Hashed ../results/final_runs//run20/config_inj.yaml: 856b103d6343b771344209e7ed8032ed
%Hashed ../results/final_runs//run20/config_pe.yaml: 451df7d803e5fb25728e376808330219
%Hashed ../results/final_runs//run20/physical_posterior.dat: 1d1f51d2d5c98d86c91f09c81a01d0ae
\newcommand{\RunTwentySLOneAmplitudeDisplacementZero}{\ensuremath{1745^{+61}_{-60}}}
\newcommand{\RunTwentySLOneBetaZero}{\ensuremath{100.0^{+1.7}_{-1.6}}}
\newcommand{\RunTwentySLOneInitialTimeZero}{\ensuremath{43179.2^{+2.5}_{-2.5}}}
\newcommand{\RunTwentyInjSLOneAmplitudeDisplacementZero}{\ensuremath{542.0}}
\newcommand{\RunTwentyInjSLOneAmplitudeDisplacementOne}{\ensuremath{1420.0}}
\newcommand{\RunTwentyInjSLOneBetaZero}{\ensuremath{40.0}}
\newcommand{\RunTwentyInjSLOneBetaOne}{\ensuremath{80.0}}
\newcommand{\RunTwentyInjSLOneInitialTimeZero}{\ensuremath{43200.0}}
\newcommand{\RunTwentyInjSLOneInitialTimeOne}{\ensuremath{43200.0}}
%File created on 07-05-2023 01:36:51
%This file is machine generated. Do *not* edit it manually if you want reproducibility.
%Hashed ../results/final_runs//run21/config_inj.yaml: 856b103d6343b771344209e7ed8032ed
%Hashed ../results/final_runs//run21/config_pe.yaml: d93c690b26193cc2f9aea8c020432ede
%Hashed ../results/final_runs//run21/physical_posterior.dat: 2176ca632d3bb92470620a0c55beea26
\newcommand{\RunTwentyOneSLOneAmplitudeDisplacementZero}{\ensuremath{551^{+152}_{-125}}}
\newcommand{\RunTwentyOneSLOneAmplitudeDisplacementOne}{\ensuremath{1406^{+116}_{-135}}}
\newcommand{\RunTwentyOneSLOneAmplitudeDisplacementTwo}{\ensuremath{101^{+280}_{-91}}}
\newcommand{\RunTwentyOneSLOneBetaZero}{\ensuremath{40.63^{+11.25}_{-9.68}}}
\newcommand{\RunTwentyOneSLOneBetaOne}{\ensuremath{79.6^{+3.2}_{-3.5}}}
\newcommand{\RunTwentyOneSLOneBetaTwo}{\ensuremath{247^{+95}_{-160}}}
\newcommand{\RunTwentyOneSLOneInitialTimeZero}{\ensuremath{43200.9^{+9.0}_{-7.3}}}
\newcommand{\RunTwentyOneSLOneInitialTimeOne}{\ensuremath{43198.71^{+14.05}_{-9.63}}}
\newcommand{\RunTwentyOneSLOneInitialTimeTwo}{\ensuremath{43118^{+109}_{-141}}}
\newcommand{\RunTwentyOneInjSLOneAmplitudeDisplacementZero}{\ensuremath{542.0}}
\newcommand{\RunTwentyOneInjSLOneAmplitudeDisplacementOne}{\ensuremath{1420.0}}
\newcommand{\RunTwentyOneInjSLOneBetaZero}{\ensuremath{40.0}}
\newcommand{\RunTwentyOneInjSLOneBetaOne}{\ensuremath{80.0}}
\newcommand{\RunTwentyOneInjSLOneInitialTimeZero}{\ensuremath{43200.0}}
\newcommand{\RunTwentyOneInjSLOneInitialTimeOne}{\ensuremath{43200.0}}
%File created on 07-05-2023 01:36:56

%This file is machine generated. Do *not* edit it manually if you want reproducibility.
%Hashed ..//run22/config_inj.yaml: 9fb96dc1273c17c7a2afb4107d4ca717
%Hashed ..//run22/config_pe.yaml: 9fb96dc1273c17c7a2afb4107d4ca717
%Hashed ..//run22/physical_posterior.dat: 6724d99be646b21131b1d52c1deb34f0
\newcommand{\RunTwentyTwoSLOneAmplitudeDisplacementZero}{\ensuremath{2481^{+64}_{-64}}}
\newcommand{\RunTwentyTwoSLOneBetaZero}{\ensuremath{20.0^{+0.9}_{-0.9}}}
\newcommand{\RunTwentyTwoSLOneInitialTimeZero}{\ensuremath{12.0000^{+0.0003}_{-0.0004}}}
\newcommand{\RunTwentyTwoInjSLOneAmplitudeDisplacementZero}{\ensuremath{2480.0}}
\newcommand{\RunTwentyTwoInjSLOneBetaZero}{\ensuremath{20.0}}
\newcommand{\RunTwentyTwoInjSLOneInitialTimeZero}{\ensuremath{12.0}}
%File created on 29-05-2023 18:12:45

%This file is machine generated. Do *not* edit it manually if you want reproducibility.
%Hashed ../results/final_runs//run23/config_inj.yaml: 9fb96dc1273c17c7a2afb4107d4ca717
%Hashed ../results/final_runs//run23/config_pe.yaml: c83abdae460ac920e0412e3a1f81e10d
%Hashed ../results/final_runs//run23/physical_posterior.dat: d778a377e84ef4074c6e247806d4f6c2
\newcommand{\RunTwentyThreeSLOneAmplitudeDisplacementZero}{\ensuremath{3643^{+449}_{-397}}}
\newcommand{\RunTwentyThreeSLOneBetaZero}{\ensuremath{154^{+14}_{-13}}}
\newcommand{\RunTwentyThreeSLOneInitialTimeZero}{\ensuremath{43270.5^{+3.2}_{-3.1}}}
\newcommand{\RunTwentyThreeInjSLOneAmplitudeDisplacementZero}{\ensuremath{2480.0}}
\newcommand{\RunTwentyThreeInjSLOneBetaZero}{\ensuremath{20.0}}
\newcommand{\RunTwentyThreeInjSLOneInitialTimeZero}{\ensuremath{43200.0}}
%File created on 07-05-2023 01:37:07
%This file is machine generated. Do *not* edit it manually if you want reproducibility.
%Hashed ../results/final_runs//run24/config_inj.yaml: 9fb96dc1273c17c7a2afb4107d4ca717
%Hashed ../results/final_runs//run24/config_pe.yaml: b60a8b177c48bdfd6f43ef98b1bdd166
%Hashed ../results/final_runs//run24/physical_posterior.dat: 67912513f495dbb33c16b90ff4376ac1
\newcommand{\RunTwentyFourSLOneAmplitudeDisplacementZero}{\ensuremath{3633^{+443}_{-386}}}
\newcommand{\RunTwentyFourSLOneBetaZero}{\ensuremath{154^{+14}_{-12}}}
\newcommand{\RunTwentyFourSLOneInitialTimeZero}{\ensuremath{43270.5^{+3.1}_{-3.1}}}
\newcommand{\RunTwentyFourInjSLOneAmplitudeDisplacementZero}{\ensuremath{2480.0}}
\newcommand{\RunTwentyFourInjSLOneBetaZero}{\ensuremath{20.0}}
\newcommand{\RunTwentyFourInjSLOneInitialTimeZero}{\ensuremath{43200.0}}
%File created on 07-05-2023 01:37:14
%This file is machine generated. Do *not* edit it manually if you want reproducibility.
%Hashed ../results/final_runs//run25/config_inj.yaml: 9fb96dc1273c17c7a2afb4107d4ca717
%Hashed ../results/final_runs//run25/config_pe.yaml: 4d00aac264cd770021a87c36cfa4adc0
%Hashed ../results/final_runs//run25/physical_posterior.dat: c4ba8735de33b5e2336428d9e98665c0
\newcommand{\RunTwentyFiveSLOneAmplitudeDisplacementZero}{\ensuremath{2439^{+40}_{-38}}}
\newcommand{\RunTwentyFiveSLOneBetaZero}{\ensuremath{20.5^{+0.6}_{-0.6}}}
\newcommand{\RunTwentyFiveSLOneInitialTimeZero}{\ensuremath{43199.3^{+0.8}_{-0.8}}}
\newcommand{\RunTwentyFiveInjSLOneAmplitudeDisplacementZero}{\ensuremath{2480.0}}
\newcommand{\RunTwentyFiveInjSLOneBetaZero}{\ensuremath{20.0}}
\newcommand{\RunTwentyFiveInjSLOneInitialTimeZero}{\ensuremath{43200.0}}
%File created on 07-05-2023 01:37:19
%This file is machine generated. Do *not* edit it manually if you want reproducibility.
%Hashed ../results/final_runs//run26/config_inj.yaml: 9fb96dc1273c17c7a2afb4107d4ca717
%Hashed ../results/final_runs//run26/config_pe.yaml: a5377fc0273140c0886c80a02559b892
%Hashed ../results/final_runs//run26/physical_posterior.dat: 432e550b83322b3f6b51cb6a5e007372
\newcommand{\RunTwentySixSLOneAmplitudeDisplacementZero}{\ensuremath{4362^{+550}_{-501}}}
\newcommand{\RunTwentySixSLOneBetaZero}{\ensuremath{172^{+15}_{-14}}}
\newcommand{\RunTwentySixSLOneInitialTimeZero}{\ensuremath{43267.7^{+3.1}_{-3.0}}}
\newcommand{\RunTwentySixInjSLOneAmplitudeDisplacementZero}{\ensuremath{2480.0}}
\newcommand{\RunTwentySixInjSLOneBetaZero}{\ensuremath{20.0}}
\newcommand{\RunTwentySixInjSLOneInitialTimeZero}{\ensuremath{43200.0}}
%File created on 07-05-2023 01:37:26
%This file is machine generated. Do *not* edit it manually if you want reproducibility.
%Hashed ../results/final_runs//run27/config_inj.yaml: 9fb96dc1273c17c7a2afb4107d4ca717
%Hashed ../results/final_runs//run27/config_pe.yaml: fb71919fcde811eedcff6327269621a1
%Hashed ../results/final_runs//run27/physical_posterior.dat: c4e92a9ffa5111c4968b3c9765564232
\newcommand{\RunTwentySevenSLOneAmplitudeDisplacementZero}{\ensuremath{4336^{+573}_{-503}}}
\newcommand{\RunTwentySevenSLOneBetaZero}{\ensuremath{174^{+16}_{-15}}}
\newcommand{\RunTwentySevenSLOneInitialTimeZero}{\ensuremath{43268.6^{+3.1}_{-3.0}}}
\newcommand{\RunTwentySevenInjSLOneAmplitudeDisplacementZero}{\ensuremath{2480.0}}
\newcommand{\RunTwentySevenInjSLOneBetaZero}{\ensuremath{20.0}}
\newcommand{\RunTwentySevenInjSLOneInitialTimeZero}{\ensuremath{43200.0}}
%File created on 07-05-2023 01:37:31

%This file is machine generated. Do *not* edit it manually if you want reproducibility.
%Hashed ..//run28/config_inj.yaml: 5fd3d18d47bdb5a08b8a209002cd7dae
%Hashed ..//run28/config_pe.yaml: cd971232262971d4e4d696055154e59b
%Hashed ..//run28/physical_posterior.dat: d52bb7d764463b28da5b1d6dc0c47096
\newcommand{\RunTwentyEightSLOneAmplitudeDisplacementZero}{\ensuremath{5021^{+1461}_{-1355}}}
\newcommand{\RunTwentyEightSLOneBetaZero}{\ensuremath{101^{+25}_{-26}}}
\newcommand{\RunTwentyEightSLOneInitialTimeZero}{\ensuremath{11.111^{+0.004}_{-0.003}}}
\newcommand{\RunTwentyEightSLTwoAmplitudeDisplacementZero}{\ensuremath{986^{+489}_{-333}}}
\newcommand{\RunTwentyEightSLTwoBetaZero}{\ensuremath{10.2^{+2.2}_{-1.9}}}
\newcommand{\RunTwentyEightSLTwoInitialTimeZero}{\ensuremath{11.111^{+0.003}_{-0.004}}}
\newcommand{\RunTwentyEightSLThreeAmplitudeDisplacementZero}{\ensuremath{4975^{+820}_{-821}}}
\newcommand{\RunTwentyEightSLThreeBetaZero}{\ensuremath{40.3^{+3.8}_{-3.5}}}
\newcommand{\RunTwentyEightSLThreeInitialTimeZero}{\ensuremath{13.889^{+0.001}_{-0.002}}}
\newcommand{\RunTwentyEightSLFourAmplitudeDisplacementZero}{\ensuremath{19822^{+3942}_{-3998}}}
\newcommand{\RunTwentyEightSLFourBetaZero}{\ensuremath{120.2^{+8.4}_{-7.6}}}
\newcommand{\RunTwentyEightSLFourInitialTimeZero}{\ensuremath{13.89^{+0.02}_{-0.02}}}
\newcommand{\RunTwentyEightInjSLOneAmplitudeDisplacementZero}{\ensuremath{5000.0}}
\newcommand{\RunTwentyEightInjSLOneBetaZero}{\ensuremath{100.0}}
\newcommand{\RunTwentyEightInjSLOneInitialTimeZero}{\ensuremath{11.111}}
\newcommand{\RunTwentyEightInjSLTwoAmplitudeDisplacementZero}{\ensuremath{1000.0}}
\newcommand{\RunTwentyEightInjSLTwoBetaZero}{\ensuremath{10.0}}
\newcommand{\RunTwentyEightInjSLTwoInitialTimeZero}{\ensuremath{11.111}}
\newcommand{\RunTwentyEightInjSLThreeAmplitudeDisplacementZero}{\ensuremath{5000.0}}
\newcommand{\RunTwentyEightInjSLThreeBetaZero}{\ensuremath{40.0}}
\newcommand{\RunTwentyEightInjSLThreeInitialTimeZero}{\ensuremath{13.89}}
\newcommand{\RunTwentyEightInjSLFourAmplitudeDisplacementZero}{\ensuremath{20000.0}}
\newcommand{\RunTwentyEightInjSLFourBetaZero}{\ensuremath{120.0}}
\newcommand{\RunTwentyEightInjSLFourInitialTimeZero}{\ensuremath{13.89}}
%File created on 29-05-2023 18:13:15

%This file is machine generated. Do *not* edit it manually if you want reproducibility.
%Hashed ..//run29/config_inj.yaml: 7b8caa3c3d549b8070330770a6febf6b
%Hashed ..//run29/config_pe.yaml: 7b8caa3c3d549b8070330770a6febf6b
%Hashed ..//run29/physical_posterior.dat: b533acc2579c86f6cfb612f06bc59eb5
\newcommand{\RunTwentyNineSLOneAmplitudeDisplacementZero}{\ensuremath{1.6^{+0.6}_{-0.4}}}
\newcommand{\RunTwentyNineSLOneBetaZero}{\ensuremath{3735^{+770}_{-543}}}
\newcommand{\RunTwentyNineSLOneInitialTimeZero}{\ensuremath{11.94^{+0.03}_{-0.03}}}
\newcommand{\RunTwentyNineSLTwoAmplitudeDisplacementZero}{\ensuremath{4.2^{+2.4}_{-1.4}}}
\newcommand{\RunTwentyNineSLTwoBetaZero}{\ensuremath{3848^{+993}_{-719}}}
\newcommand{\RunTwentyNineSLTwoInitialTimeZero}{\ensuremath{36.93^{+0.04}_{-0.04}}}
\newcommand{\RunTwentyNineInjSLOneAmplitudeDisplacementZero}{\ensuremath{1.48}}
\newcommand{\RunTwentyNineInjSLOneBetaZero}{\ensuremath{3600.0}}
\newcommand{\RunTwentyNineInjSLOneInitialTimeZero}{\ensuremath{11.94}}
\newcommand{\RunTwentyNineInjSLTwoAmplitudeDisplacementZero}{\ensuremath{3.72}}
\newcommand{\RunTwentyNineInjSLTwoBetaZero}{\ensuremath{3600.0}}
\newcommand{\RunTwentyNineInjSLTwoInitialTimeZero}{\ensuremath{36.94}}
%File created on 29-05-2023 18:08:56

%This file is machine generated. Do *not* edit it manually if you want reproducibility.
%Hashed ..//run30/config_inj.yaml: e65be7c6d665fa9d2895f547ce8b3a6e
%Hashed ..//run30/config_pe.yaml: e65be7c6d665fa9d2895f547ce8b3a6e
%Hashed ..//run30/physical_posterior.dat: 139716705ac22b0f0e88a67443030479
\newcommand{\RunThirtyLPFOneDeltavZero}{\ensuremath{0.300^{+0.004}_{-0.004}}}
\newcommand{\RunThirtyLPFOneInitialTimeZero}{\ensuremath{12.001^{+0.003}_{-0.003}}}
\newcommand{\RunThirtyLPFOneTauFallZero}{\ensuremath{20^{+13}_{-17}}}
\newcommand{\RunThirtyLPFOneTauRiseZero}{\ensuremath{20^{+13}_{-17}}}
\newcommand{\RunThirtyInjLPFOneDeltavZero}{\ensuremath{0.3}}
\newcommand{\RunThirtyInjLPFOneInitialTimeZero}{\ensuremath{12.0}}
\newcommand{\RunThirtyInjLPFOneTauFallZero}{\ensuremath{21.0}}
\newcommand{\RunThirtyInjLPFOneTauRiseZero}{\ensuremath{20.0}}
%File created on 29-05-2023 18:10:02

%This file is machine generated. Do *not* edit it manually if you want reproducibility.
%Hashed ..//run31/config_inj.yaml: 4156387a95682d4a95e62fb180577932
%Hashed ..//run31/config_pe.yaml: 4156387a95682d4a95e62fb180577932
%Hashed ..//run31/physical_posterior.dat: 339cde4f7d98fddb5e1f28762764d59f
\newcommand{\RunThirtyOneLPFOneDeltavZero}{\ensuremath{2.00^{+0.02}_{-0.02}}}
\newcommand{\RunThirtyOneLPFOneInitialTimeZero}{\ensuremath{12.000^{+0.002}_{-0.002}}}
\newcommand{\RunThirtyOneLPFOneTauFallZero}{\ensuremath{439^{+485}_{-55}}}
\newcommand{\RunThirtyOneLPFOneTauRiseZero}{\ensuremath{848^{+75}_{-465}}}
\newcommand{\RunThirtyOneInjLPFOneDeltavZero}{\ensuremath{2.0}}
\newcommand{\RunThirtyOneInjLPFOneInitialTimeZero}{\ensuremath{12.0}}
\newcommand{\RunThirtyOneInjLPFOneTauFallZero}{\ensuremath{900.0}}
\newcommand{\RunThirtyOneInjLPFOneTauRiseZero}{\ensuremath{400.0}}
%File created on 29-05-2023 18:11:23

%This file is machine generated. Do *not* edit it manually if you want reproducibility.
%Hashed ..//run32/config_inj.yaml: 2eb10bcd31cfb4747fddb16e881a6e1a
%Hashed ..//run32/config_pe.yaml: 2eb10bcd31cfb4747fddb16e881a6e1a
%Hashed ..//run32/physical_posterior.dat: 5d282569e21815296006352babe604a2
\newcommand{\RunThirtyTwoLPFOneDeltavZero}{\ensuremath{102^{+5}_{-4}}}
\newcommand{\RunThirtyTwoLPFOneInitialTimeZero}{\ensuremath{12.000^{+0.002}_{-0.002}}}
\newcommand{\RunThirtyTwoLPFOneTauFallZero}{\ensuremath{7453^{+2211}_{-1417}}}
\newcommand{\RunThirtyTwoLPFOneTauRiseZero}{\ensuremath{7453^{+2201}_{-1402}}}
\newcommand{\RunThirtyTwoInjLPFOneDeltavZero}{\ensuremath{100.0}}
\newcommand{\RunThirtyTwoInjLPFOneInitialTimeZero}{\ensuremath{12.0}}
\newcommand{\RunThirtyTwoInjLPFOneTauFallZero}{\ensuremath{7500.0}}
\newcommand{\RunThirtyTwoInjLPFOneTauRiseZero}{\ensuremath{7400.0}}
%File created on 29-05-2023 18:12:00
\begin{abstract}

Detecting and coherently characterizing thousands of gravitational-wave signals is a core data-analysis challenge for the Laser Interferometer Space Antenna (LISA).
Transient artifacts, or ``glitches'', with disparate morphologies are expected to be present in the data, potentially affecting the scientific return of the mission. 
We present the first joint reconstruction of short-lived astrophysical signals and noise artifacts.
Our analysis is inspired by glitches observed by the LISA Pathfinder mission, including both acceleration and fast displacement transients.
We perform full Bayesian inference using LISA time-delay interferometric data and gravitational waveforms describing mergers of massive black holes. 
We focus on a representative binary with a detector-frame total mass of $6 \times 10^7 M_\odot$ at redshift $5$, yielding a signal lasting $\sim 30~\mathrm{h}$ in the LISA sensitivity band.
We explore two glitch models of different flexibility, namely a fixed parametric family and a shapelet decomposition. 
In the most challenging scenario, we report a complete loss of the gravitational-wave signal if the glitch is ignored; more modest glitches induce biases on the black-hole parameters. 
On the other hand, a joint inference approach fully sanitizes the reconstruction of both the astrophysical and the glitch signal. 
We also inject a variety of glitch morphologies in isolation, without a superimposed gravitational signal, and show we can identify the correct transient model. Our analysis is an important stepping stone toward a realistic treatment of LISA data in the context of the highly sought-after ``global fit''. 

\end{abstract}

\maketitle

\section{\label{sec:intro} Introduction}

The Laser Interferometer Space Antenna (LISA) \cite{2017arXiv170200786A}, currently planned to be launched in the early 2030s, will detect gravitational waves (GWs) from space.
LISA will extend the exploration of the GW spectrum in the milliHertz band -- from about $10^{-4}$ to $1$ \rm{Hz} -- providing observations of astrophysical sources ranging from Galactic white-dwarf binaries to mergers of massive black-holes at high redshift~\cite{2023LRR....26....2A,2022arXiv220405434A}.

The detection and characterization of different astrophysical sources is an extremely challenging  problem of data-analysis. This is due to the combined effect of the all-sky detector sensitivity and the large number, $\mathcal{O}$(10$^4$), of long-lived GW signals overlapping both in time and frequency. 
Maximizing the payoff of the LISA mission requires an accurate, efficient, and global analysis \cite{2005PhRvD..72d3005C,2023PhRvD.107f3004L}, simultaneously fitting data models for an unknown number of detectable GW sources and uncertain detector noise.

In addition to the abundance of astrophysical sources, the LISA data stream will be polluted by noise transients.
These artifacts, also called ``glitches'' from a terminology borrowed from ground-based detectors, have been observed at a rate of about one per day and extensively characterized by the LISA Pathfinder (LPF) mission \cite{2016PhRvL.116w1101A,2018PhRvL.120f1101A}. 
Efforts are ongoing to understand the origin of the LPF glitches by capitalizing on the collected data and eliminating them by design in the LISA hardware. 
Previous studies stressed the need to assess their impact on the scientific return of the LISA mission~\cite{2022PhRvD.106f2001A,2022PhRvD.105d2002B}. The physical nature of glitches in LPF still needs to be fully understood, with possible interpretations including outgassing phenomena, electronics events, and eddy current transients~\cite{2022PhRvD.106f2001A}. 
Moreover, new types of unexpected noise artifacts can appear in LISA because of the increased complexity of both spacecraft and payload design compared to LPF.
Because the occurrence and morphology of glitches in the full LISA setup are uncertain, a conservative approach is to prepare a robust data analysis strategy to mitigate their impact downstream.  

Tackling the fundamental challenge of including glitches in parameter-estimation pipelines is well recognized by the LISA Consortium as part of the core preparation activities for the imminent mission adoption.
To this end, a set of LISA Data Challenges (LDCs)~\cite{ldcweb} are in progress to develop and demonstrate data-analysis readiness.
Among others, the LDC nicknamed \textit{Spritz} is devoted to investigating glitches and gaps in the reconstruction of signals from massive black-hole binaries (MBHBs).
Approaches to the GW bursts detection in presence of glitches have been explored in the past both for LISA~\cite{2019PhRvD..99b4019R} and ground-based detectors~\cite{2021PhRvD.103d4013C}.

A recent analysis suggests the adoption of heavy-tailed likelihoods to mitigate the effect of noise transients upon the inference of GW sources~\cite{2023arXiv230504709S}.
In this work, we instead assess for the first time the impact of glitches on short-lived MBHB signals  performing direct, joint parameter estimation. 
We present a complete analytical derivation of the LISA response to two types of instrumental artifacts as detected by LPF, namely  force and displacement transients of the test masses.
We then report results by including both models in a large, multi-source parameter estimation framework for LISA data analysis.
This infrastructure, called \textsc{Balrog}, is currently under active development and has already been tested against different astrophysical sources (see e.g. Refs.~\cite{2020ApJ...894L..15R,2021PhRvD.104d4065B,2022arXiv220403423K,2022arXiv221010812F,2022arXiv221202572P}). 

The paper is organized as follows.
In Sec.~\ref{sec:exper}, we introduce the phenomenology of the expected instrumental artifacts.
In Sec.~\ref{sec:model}, we present our glitch models and provide a brief summary of the fiducial GW-source and glitch parameters.
In Sec.~\ref{sec:glitch-tdi}, we derive an alternative set of time-delay interferometric (TDI) variables suitable for the simultaneous treatment of glitches and GW signals.
In Sec.~\ref{sec:inference}, we provide definitions of relevant statistical quantities and details on our parameter-estimation runs.
In Sec.~\ref{sec:results}, we present our inference results.
Finally, in Sec.~\ref{sec:conclusions}, we summarize our findings and describe future developments.
Throughout this paper, we use units where $c=1$.

\section[LISA Pathfinder]{LPF glitches in LISA data}
\label{sec:exper}

\subsection[LPF glitches]{\label{subsec:LPF_phen} Phenomenology of LPF glitches}

Glitches are
observed as additional signals in the data stream. They can be thus modeled and subtracted from the data as such.
The strategy here is to (i) get a consistent estimate of the power 
spectral density (PSD) of the underlying quasi-stationary noise over the entire 
data stream and thus (ii) improve the astrophysical signal inference by making it robust against glitch-induced biases.
The latter constitutes a key element of the LISA data processing pipeline in view of the targeted ``global fit''~\cite{2005PhRvD..72d3005C,2023PhRvD.107f3004L}.
The properties of glitches, namely amplitude, duration, and time morphology, depend both on the measurement system and the originating physical process.

LPF observed two main kinds of glitches: a first class treated as an effective displacement-measurement artifact in the optical metrology chain and another class due to spurious forces acting on the test masses (TMs). 
Displacement glitches have been rarely observed in nominal conditions, have a typical duration comparable with the LISA sampling cadence, and carry negligible impulse per unit of mass as compared to the typical forces acting on the TMs~\cite{2022PhRvD.106f2001A}.
As a consequence, fast, low-impulse glitches could be expected to affect the geodesic motion of the LISA constellation only mildly.
On the contrary, force events result in impulse-carrying glitches lasting from tens of seconds to several hours, have a significant impact on the noise performance, and can  potentially contaminate GW detection and parameter estimation. 

During its ordinary runs, LPF observed 102 impulse-carrying glitches and 81 of these were visible in the data stream as a sharp, positive offset of the residual force-per-unit-mass (henceforth loosely referred to as ``acceleration'')~\cite{2022PhRvD.106f2001A}. 
These acceleration glitches correspond to the two TMs moving toward each other along the sensitive axis of the pair, i.e. the direction joining their respective centers of mass.
The rate of these events has been estimated to be about 1 per day and compatible with a   Poisson distribution~\cite{2022PhRvD.106f2001A}. Several possible physical origins for glitches have been vetoed by extensive cross-checking and correlation analysis on LPF data, with the most plausible explanation pointing to either gas outbursts or virtual leaks in  the vacuum chamber and the material surrounding the TMs. Dedicated experimental studies are underway to corroborate this hypothesis~\cite{2022PhRvD.106f2001A}.

\subsection[Differential TM acceleration]{Guiding principles for LISA differential acceleration measurements}\label{subsec:diff-accel}

We now list a few guiding principles behind our modeling 
choices: 
\begin{itemize}
\item Long-lived glitches related to force phenomena such as those observed by LPF are the most relevant for LISA;
For these, we adopt a phenomenological parameterization suitable to describe their temporal evolution in terms of differential test-mass accelerations;
\item Constructing the corresponding signal model for fractional phase observables in the frequency domain is more complex, although doable;
\item Long-lived transients present in a displacement (optical phase) or velocity (optical frequency) observable disappear in an acceleration observable, with the signal disturbance limited to the duration of the external force transient. Likewise, glitch parameters related to the initial conditions -- position and velocity -- are eliminated with an acceleration observable;
\item In a realistic operational setup, systematic errors arising from force disturbances (e.g. stiffness coupling) 
could be subtracted directly in acceleration. 
Thus, our fitting model does not require any additional integration or whitening filter;
\item When the effective glitch ``signal'' has spectral content mainly near the low-frequency end of the LISA sensitivity range, differentiation is numerically safer than integration. 
In this regime, data correction from systematics in the displacement variables is still viable;
\item The corresponding TDI variables written in acceleration allow for a straightforward inclusion of LPF glitches in a Bayesian inference framework;
\item GW signal models can be easily rewritten as effective accelerations 
by differentiating those already available in phase or fractional frequency.
\end{itemize}

\begin{figure*}[!ht]
    \centering
   \includegraphics[width=1.5\columnwidth, keepaspectratio]{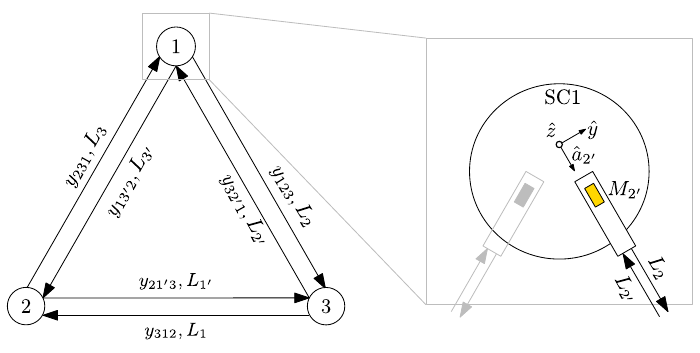}
    \caption[Schematics of single laser links and glitch 
    reference system conventions.]{
    Schematics of single laser links and glitch reference system conventions. The constellation is made of three satellites (white circles), each housing two TMs (right inset, yellow and gray boxes). 
    Each satellite is connected to the other two by four links, two for each TM. 
    Signals denoted by $y_{ijk}$ or $y_{ij^\prime k}$ are emitted by the $i$-th satellite, received by the $k$-th satellite, therefore traveling
    along either $L_j$ or $L_{j^\prime}$.
    The indexes $j$ and $j^\prime$ are used to denote cyclic and anti-cyclic permutations of 
    $123$, respectively. Unit vectors $\hat{a}_j$ parametrize the glitch component 
    along the incoming (outgoing) link $L_{j^\prime}$ ($L_{j}$) associated with the test mass $M_{j^\prime}$.
  On satellite $1$ a generic acceleration glitches acting on test mass $M_{2^\prime}$ and $M_{3}$ are described by the components $a_{2^\prime}$ and $a_3$, respectively. 
    The former [latter] affects link $y_{32^\prime 1}(t)$ [$y_{231}(t)$] at reception and link $y_{123}(t-L)$ [$y_{13^\prime 2}(t-L)$]  at emission.
    }
    \label{fig:TDI}
\end{figure*}

These broad considerations are mostly inspired by the observational equivalence between GWs and tidal forces accelerating TMs relative to their local inertial frames~\cite{PhysRevD.88.082003}.
We thus opt to implement our joint inference for glitches and GWs with suitable acceleration TDI variables.

\section{Transients modeling}
\label{sec:model}

The fundamental observable in LISA is the phase evolution $\Delta\phi$ of a one-way propagating laser along each of the six links connecting the satellites. This can be equivalently written as an optical pathlength
\begin{equation}
    L=\frac{\Delta\phi}{\omega_l}\,, \label{eq:phase-vs-L}
\end{equation}
where $\omega_l$ is the central frequency of the laser signal, which is assumed to be constant.

We now focus on three different mechanisms perturbing the phase 
readout.

\subsection{\label{subsec:acc-model} 
Acceleration transients
}

The two TMs housed in each of the LISA satellites are expected to independently exchange momentum with their surrounding environment (see  Fig.~\ref{fig:TDI} for a schematic representation). 
We model the resulting transient acceleration profile $\vec{a}_i$ of the $i$-th test mass as in Ref.~\cite{2022PhRvD.105d2002B}.
We use a two-damped exponential model inspired by glitches observed in LPF, namely 
\begin{equation}
g(t;A,\beta_1,\beta_2,\tau)\!=\!\! \frac{A}{\beta_1 -\beta_2}\!\left(e^{-\frac{t-\tau}{\beta_1}}-e^{-\frac{t-\tau}{\beta_2}}\right)\!\Theta(t\!-\!\tau),
\label{eq:A1-TD}
\end{equation}
which we refer to as \textsc{Model A1}.
Equation~\eqref{eq:A1-TD} integrates to the net transferred momentum per unit mass:
\begin{equation}    
\int_{-\infty}^{+\infty}  g(t;A,\beta_1,\beta_2,\tau) \; \mathrm{d}t = A\,. \label{eq:lpf-normalization}
\end{equation}
The parameters $\beta_1,\beta_2$ describe the typical timescales of the two exponentials while $\tau$ is the glitch onset time entering the Heaviside step function $\Theta$.
The corresponding Fourier-domain representation is
\begin{equation}\label{eq:A1-FD}
\Tilde{g}(\omega;A,\beta_1,\beta_2,\tau) = -A\frac{ e^{-i \tau  \omega }}{(\beta_1\omega -i) (\beta_2\omega -i)} \,.
\end{equation}

Accommodating glitches of unknown shape requires a more flexible model. We construct this using a superposition of $S$ Gabor-Morlet shapelets 
\begin{align}
g(t) &=
\sum_i^{S} \sigma\left(t ; A_i, \tau_i, \beta_i, n_i\right), \label{eq:decomposition}
\end{align}
where 
\begin{align}
\sigma\left(t ; A, \tau, \beta, n\right)  &=
c_n \psi_{n}\left(\frac{t-\tau}{\beta}\right),  \label{eq:shapemodel}
\\
\psi_{n}\left(t\right) &= \frac{2t}{n} e^{-t/n} L_{n-1}^{(1)}\left(\frac{2t}{n}\right)\Theta\left(t\right),\\
c_n &= (-1)^{n-1} \frac{A}{2\beta n^2},
\end{align}
and $L_{n}^{(\alpha)}(t)$ is the $n-$th generalized Laguerre polynomial~\cite{1965hmfw.book.....A}.  We refer to these expressions as \textsc{Model A2}.
Comparing to Ref.~\cite{2022PhRvD.105d2002B}, we use a different normalization $c_n$ for the individual shapelets such that 
\begin{equation}    
\int_{-\infty}^{+\infty}  \sigma(t;A,\tau,\beta,n) \, \mathrm{d}t = A\,, \qquad \forall n \in \mathbb{N}\,. \label{eq:normalization}
\end{equation}
In the frequency domain Eq.~\eqref{eq:shapemodel} reads
\begin{equation}\label{eq:A2-FD}
    \Tilde{\sigma}(\omega; A,\tau,\beta,n) = (-1)^n e^{-i \omega \tau}A \frac{(n \beta  \omega + i )^{n-1} }{(n \beta
    \omega -i)^{n+1}} .
\end{equation}
Shapelets in this parametric family are quasi-orthogonal, i.e.
 \begin{align}
 &\int_{-\infty}^{+\infty} \Tilde{\sigma}(\omega; A,\tau,\beta,n)  
 \Tilde{\sigma}^*(\omega; A^\prime,\tau,\beta, m) \, \mathrm{d}\omega
 = \delta_{nm}\frac{\pi A A^\prime}{2n\beta} \,, \label{eq:quasi-orthogonal1}
 \\
& \int_{-\infty}^{+\infty} \Tilde{\sigma}(\omega; A,\tau,\beta,n)  
 \Tilde{\sigma}^*(\omega; A^\prime,\tau^\prime,\beta,n)\mathrm{d}\omega \nonumber\\
 \, &\qquad= \frac{\pi A A^\prime}{2n^2\beta^2} e^{-\frac{\left|\tau -\tau^\prime\right|}{n\beta}}(n\beta+\left|\tau -\tau^\prime\right|) .\label{eq:quasi-orthogonal2}
 \end{align}

From Eqs.~\eqref{eq:A1-FD} and~\eqref{eq:A2-FD} it is immediate to show that \textsc{Model A1} tends to \textsc{Model A2} with $n=1$  in the limit where  $\beta_1\to\beta_2$.

\subsection[OMS Transients]{\label{subsec:displ-model} Displacement transients}

The interferometer readout system is also expected to generate transient phase fluctuations. 
From Eq.~\eqref{eq:phase-vs-L}, we model these as effective displacement transients with the same agnostic shapelet parameterization used in Eq.~\eqref{eq:decomposition}. We use a superposition of $S$ shapelets
\begin{equation}\label{eq:D-TD}
\Delta L(t) = \sum_i^{S} \sigma\left(t ; D_i, \tau_i, \beta_i, n_i\right) ,
\end{equation}
where
\begin{equation}    
\int_{-\infty}^{+\infty} \mathrm{d}t \Delta L(t) = \sum_i^S D_i \label{eq:normalizationD}
\end{equation}
is the net integrated displacement experienced by the test mass before returning asymptotically to its free-fall condition. We refer to this parametric family of glitches as \textsc{Model D}.
The frequency domain representation follows from Eq.~\eqref{eq:A2-FD} and reads
\begin{equation}\label{eq:D-FD}
    \Tilde{\sigma}(\omega; D,\tau,\beta,n) = (-1)^n e^{-i \omega \tau}D \frac{(n \beta  \omega + i )^{n-1} }{(n \beta
    \omega -i)^{n+1}} .
\end{equation}

\subsection[GW Transients]{\label{subsec:gw-model} GW transients}

Among the large variety of typical sources populating the LISA sensitivity band, the most massive binary systems detectable produce hours to years-long transient signals. To leading-order, the binary time to merger $t_m$ from a reference frequency $f_\mathrm{ref}$ is~\cite{1963PhRv..131..435P,2014LRR....17....2B}
\begin{equation}
    t_m\sim \left(\frac{3}{4\eta}\right) \left(\frac{f_\mathrm{ref}}{0.1~\mathrm{mHz}}\right)^{-\frac{8}{3}} \left(\frac{M_{\mathrm{z}}}{10^{7}M_{\odot}}\right)^{-\frac{5}{3}}\mathrm{days}\,,
    \label{eq:time_to_coalescence}
\end{equation}
where
$\eta\equiv m_{1}m_{2}/(m_1 + m_2)^{2}$ is the symmetric mass ratio
and $M_z= (1+z) (m_1 + m_2)$ is the solar-system barycenter frame total mass for a source of component masses $m_{1}$ and $m_2$.
By contrast, glitches observed by LPF have typical durations of seconds to hours and are  positively correlated with the transferred momentum per unit mass ranging from $10^{-2}$ to $10^{3}\,\mathrm{pm/s}$~\cite{2022PhRvD.106f2001A}. 
Their broadband, short-lived morphology makes 
them the most likely to impact parameter estimation for GW transient sources of comparable duration.

\begin{figure}[!ht]
    \centering
   \includegraphics[width=0.985\columnwidth, keepaspectratio]{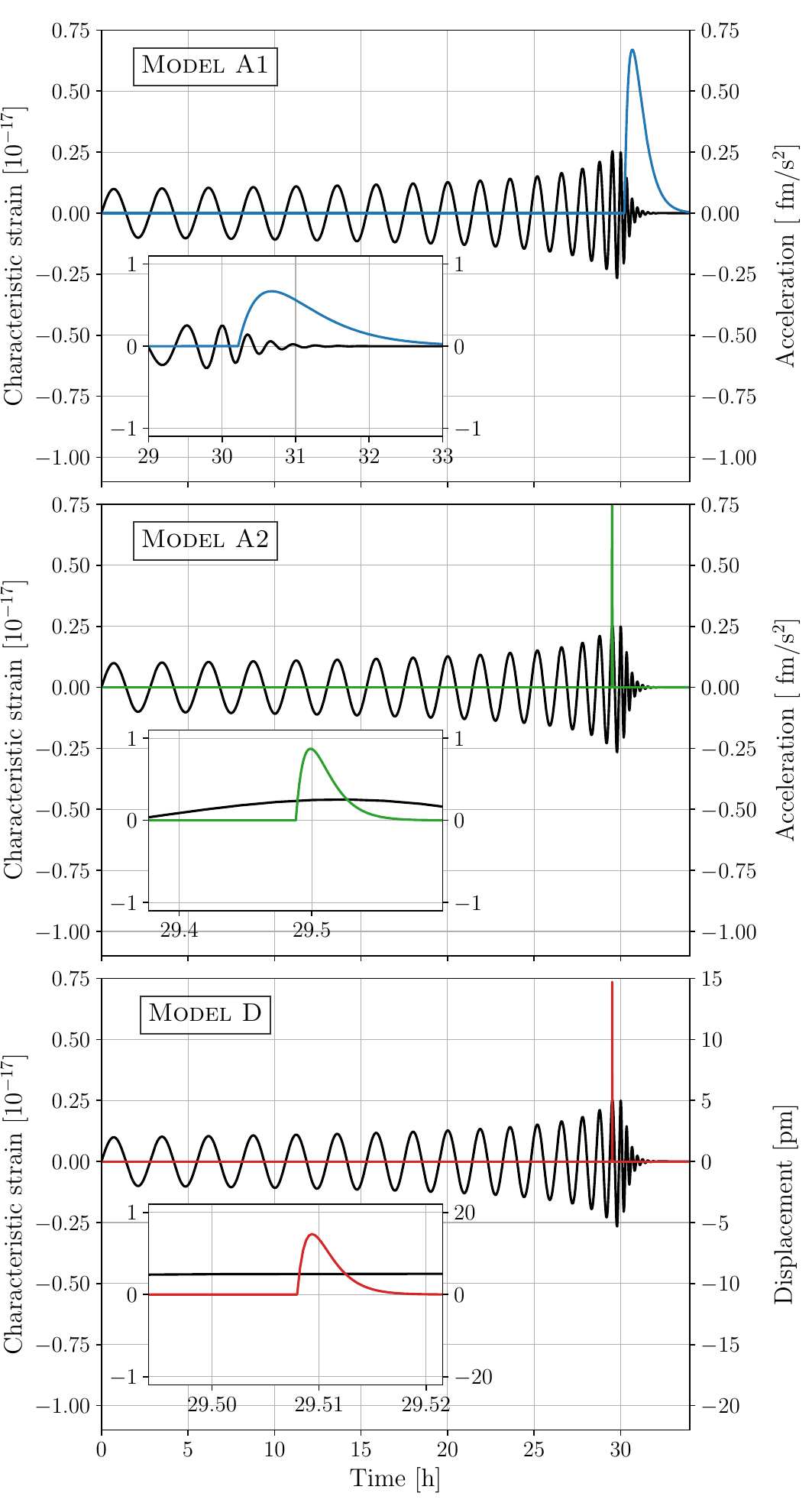}
    \caption{
    Fiducial waveforms for our parameter-estimation runs. 
    Black solid curves show the  MBHB signal we consider ($M_z = 6\times10^7\,M_\odot$ and $z = 5$), which is identical across the three panels. 
   Colored curves in the top, middle, and bottom panels describe the \textsc{Model A1}, \textsc{Model A2}, \textsc{Model D} glitch amplitudes, respectively. 
    Signals shown in the three panels correspond to injections in runs 9, 10, and 11 and exemplify glitches lasting hours, minutes, and seconds, respectively (cf. Table~\ref{tab:summary-runs}). 
    The parameters of the injected signals
   are shown in Tables~\ref{tab:glitchGW2}, \ref{tab:glitchGW1}, and \ref{tab:glitchGW3}.}
    \label{fig:Waveform}
\end{figure}

\begin{figure}[t]
    \centering
   \includegraphics[width=\columnwidth, keepaspectratio]{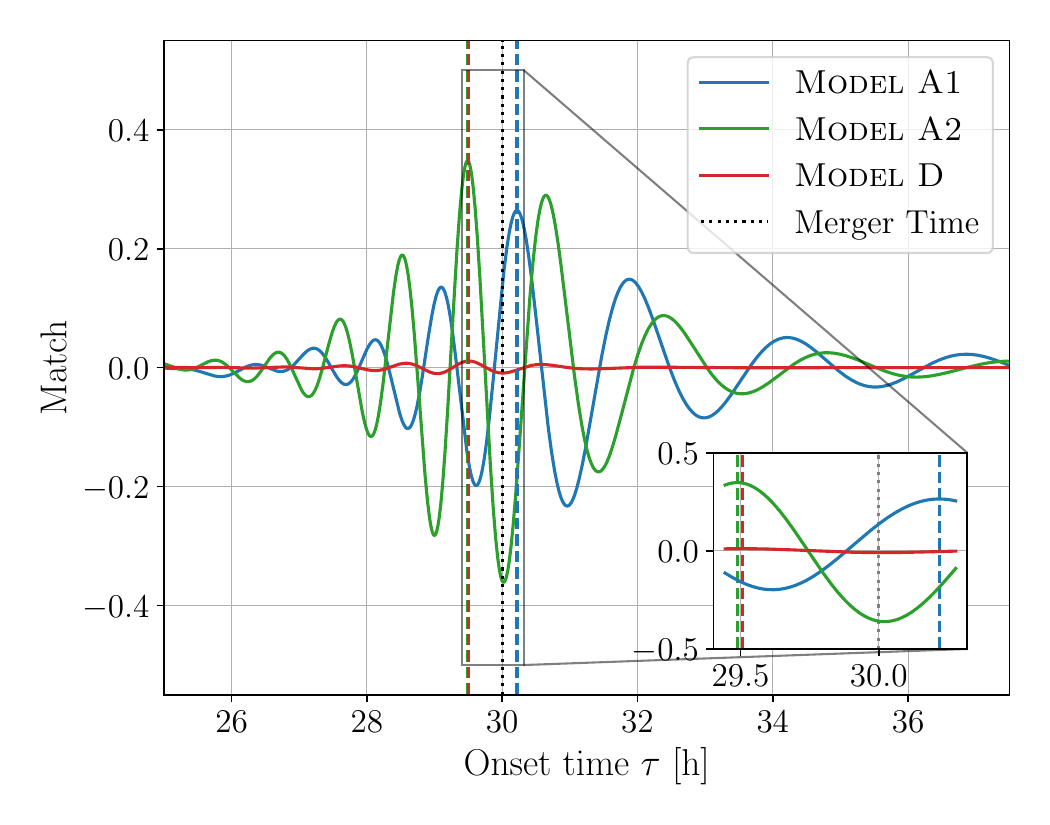}
    \caption{Match between the GW and glitch signals as a function of onset time. 
    The blue, green, and red solid curve considers to the \textsc{Model A1} (\textsc{A2}, \textsc{D}) glitch shown in Fig.~\ref{fig:Waveform}. The GW signal is fixed to that of our fiducial MBHB.
    Dashed vertical lines with matching color denote the onset time that maximizes the match.
    The black dotted line denotes the GW source nominal merger time. 
    The inset shows a 40-minute interval zoom-in around the merger and glitch onset times. 
    }
    \label{fig:MergerTime}
\end{figure}

We select three fiducial noise transients and superimpose them on a short-lived ($t_m = 30$ hours) high-mass ($M_z = 6\times10^{7}\,M_\odot, \eta=3/16$) MBHB at redshift $z = 5$.
We assume zero sensitivity below $0.1\,\mathrm{mHz}$~\cite{2022arXiv221202572P}.
We consider a short-duration \textsc{Model D} glitch ($\beta = 5\,\rm{s} $), a moderate-duration \textsc{Model A2} ($\beta = 40\,\rm{s}$), and a long-duration \textsc{Model A1} glitch with $\beta_1 +\beta_2 = 3300\,\rm{s}$. 
All three glitches have peak amplitudes close to the merger time of the GW source, as shown in Fig.~\ref{fig:Waveform}.
For a conservative approach, we fine-tune the glitch onset times to maximally impact the reconstruction of GW source parameters. This is done by maximizing the match between the glitch and GW waveforms as shown in Fig.~\ref{fig:MergerTime} (see Sec.~\ref{sec:inference} for more details).

We model the GW signal with the \textsc{IMRPhenomXHM}~\cite{2020PhRvD.102f4002G, 2020PhRvD.102f4001P}   waveform approximant
which captures the full coalescence of a quasi-circular, non-precessing black-hole binary.
The implementation of the LISA response to this GW signal in the \textsc{Balrog} code has been presented in Ref.~\cite{2022arXiv221202572P}.

We choose to parametrize the GW signal injected as follows:
$m_{1z,2z}$ and $\chi_{1,2}$ denote the binary component redshifted masses and aligned dimensionless spin, respectively;
$t_m, \phi_0, \psi$ denote the time to merger introduced in Eq.~\eqref{eq:time_to_coalescence}, initial phase and polarization, respectively;
$\sin\beta, \lambda$ denote the (sine-)ecliptic latitude and longitude;
$d_L$ and $\iota$ denote the source luminosity distance and inclination.
Tables~\ref{tab:glitchGW2}, \ref{tab:glitchGW1} and \ref{tab:glitchGW3} list the parameter values of our fiducial GW source which has an $\textrm{SNR}$ of 187 and is common across all of our runs.

\section{\label{sec:glitch-tdi} Acceleration TDIs}
 
We use Eqs.~\eqref{eq:A1-FD},~\eqref{eq:A2-FD}, and \eqref{eq:D-FD} to model the TDI variables~\cite{2021LRR....24....1T} $\tilde{s}_k(f;\boldsymbol{\theta})$ entering the likelihood, cf. Sec.~\ref{sec:inference}.
We  work in the constant equal-armlength approximation and label the three TDI variables $M_X, M_Y$, and $M_Z$, respectively. 
In this approximation, one needs a single time-delay operator $\cal{D}$
\begin{equation}\label{eq:1g-delay}
\mathcal{D} \left[f(t)\right] = f(t-L) .
\end{equation}

This is applied to the single-link phase measurements $y_{ijk}$.
Signals denoted by $y_{ijk}$ or $y_{ij^\prime k}$ are emitted by the $i$-th satellite, received by the $k$-th satellite, therefore traveling
along either $L_j$ or $L_{j^\prime}$ (see Fig.~\ref{fig:TDI} for a schematic representation).
The indexes $j$ and $j^\prime$ are used to denote cyclic and anti-cyclic permutations of 
$123$, respectively.
We thus to obtain the TDI variables
\begin{align}
    M_X &=  y_{231} + \mathcal{D}y_{13^\prime2} - y_{32^\prime1} -\mathcal{D} y_{123} ,\label{eq:Mx}\\
    M_Y &= y_{312} + \mathcal{D}y_{32^\prime1} - y_{21^\prime3} -\mathcal{D} y_{231} ,\label{eq:My}\\
    M_Z &=  y_{123} + \mathcal{D}y_{21^\prime3} - y_{13^\prime2} -\mathcal{D} y_{231} .\label{eq:Mz}
\end{align} 

Incorporating \textsc{Model A1} and \textsc{Model A2} signals into Eqs.~\eqref{eq:Mx}, \eqref{eq:My}, and \eqref{eq:Mz} requires integrating the single-link differential accelerations twice.
However, any non-zero total transferred momentum necessitate artificial regularization or ad-hoc approximations to construct a Fourier-domain representation of the signal.
We solve this problem by introducing a set of ``acceleration TDIs'' $G_{X,Y,Z}$ which are trivially related to Eqs.~\eqref{eq:Mx}, \eqref{eq:My}, and \eqref{eq:Mz} by double differentiation.
In the frequency domain one has
\begin{align}
    \mathcal{F}\left[G_X\right] &= (2 \pi f)^2 \mathcal{F}\left[M_X\right] \label{eq:GxfromMx}\\
    G_X &=  g_{231} + \mathcal{D}g_{13^\prime2} - g_{32^\prime1} -\mathcal{D} g_{123} ,\label{eq:Gxfromgijk}
\end{align}
where $\mathcal{F}$ denotes the Fourier transform operator and 
\begin{equation}
g_{ijk}(t) = \frac{d}{dt^2} \left[y_{ijk}(t)\right] . \label{eq:gijkFromyijk}
\end{equation}
Similar definitions hold for $G_Y, G_Z$ upon cyclic permutation of indices.

The key advantage of introducing a new set of TDIs lies in its instrumental robustness. Equation~\eqref{eq:GxfromMx} also allows us to conveniently recycles signal models available in fractional displacement by including both \textsc{Model D} glitches and GW signals. Furthermore, Eq.~\eqref{eq:gijkFromyijk} does not require a transfer function to model acceleration glitches.

Following the conventions shown in Fig.~\ref{fig:TDI}, the single-link perturbation $g_{ijk}(t)$ is obtained from the instantaneous accelerations $\vec{g}_i(t)$ and $\vec{g}_k(t-L)$ which are experienced by sender $i$ and receiver $k$ along the link $j$, and projected along the unit-vectors $\hat{a}_j(t-L)$ and $\hat{a}_{j^\prime}(t)$, respectively.
We associate a unit vector $\hat{a}_j$ to each test mass $M_j$ pointing in the direction opposite to $L_j$. 
For simplicity, we denote the associated vector components $a_j$. 
Given the choice of the local reference system, a positive value $a_i$ corresponds to a negative displacement $\Delta L_i$.
The three TDI observables in terms of the individual test mass accelerations are
\begin{align}
        G_X &= (1+\mathcal{D}^2)(a_{2^\prime} - a_3)
            + 2\mathcal{D}(a_2 - a_{3^\prime}),
            \label{eq:G_Xfromaccelerations}\\
        G_Y &= (1+\mathcal{D}^2)(a_{3^\prime} - a_1)
            + 2\mathcal{D}(a_3 - a_{1^\prime}),
            \label{eq:G_Yfromaccelerations}\\
        G_Z &= (1+\mathcal{D}^2)(a_{1^\prime} - a_2)
            + 2\mathcal{D}(a_1 - a_{2^\prime}) .
            \label{eq:G_Zfromaccelerations}
\end{align}

It is importante to note how the acceleration TDI variable $G_X$ ($G_Y$, $G_Z$) is insensitive to glitches acting on links $L_1$ and $L_1^\prime$ ($L_2$ and $L_2^\prime$, $L_3$ and $L_3^\prime$). 
This would no longer be true if a single glitch affects more than one TM (or more optical phase measurements); further modeling on this point will be presented elsewhere. Following the standard procedure \cite{2021LRR....24....1T}, we combine $G_X$, $G_Y,$ and $G_Z$ into three noise-orthogonal variables
\begin{align}
    G_A & = \dfrac{G_Z - G_X}{\sqrt{2}}, \label{eq:GA}\\ 
    G_E & = \dfrac{G_X -2G_Y +G_Z}{\sqrt{6}}, \label{eq:GE}\\
    G_T & = \dfrac{G_X+G_Y+G_Z}{\sqrt{3}}. \label{eq:GT}
\end{align}
Equations~\eqref{eq:GA}, \eqref{eq:GE}, and \eqref{eq:GT} define the data pieces entering our inference pipeline. 

\begin{table*}[tbh]
    \centering
    \begin{tabular}
    {c | c | c | c | c | c | c | c}
    ID & \multicolumn{2}{c|}{Injection} & \multicolumn{2}{c|}{Recovery} & $\log_{10}\mathcal{Z}$  & Figure & Table\\
         \hline
          & Glitch Model & GW Source & Glitch Model & GW Source& & & \\
         \hline
    \rowcolor{teal!20}1 &\xmark & \cmark (acceleration TDI) & \xmark & \cmark (acceleration TDI) & $-35.27$ & & \\
    \rowcolor{teal!20}2 &\xmark & \cmark (displacement TDI) & \xmark & \cmark (displacement TDI) & $-35.19$ & & \\
    \rowcolor{teal!20}3 &\textsc{A1} & \xmark & \textsc{A1} & \xmark & $-14.0$ &  & \\
    \rowcolor{teal!20}4 &\textsc{A2} & \xmark & \textsc{A2} & \xmark & $-9.1$ &  & \\
    \rowcolor{teal!20}5 &\textsc{D} & \xmark & \textsc{D} & \xmark & $-8.8$ &  & \\
         6 & \textsc{A1} & \cmark & \xmark & \cmark & $-14537.8$ &  \ref{fig:cornerMBHB+LPF} & \ref{tab:glitchGW2}\\
         7 &\textsc{A2} & \cmark & \xmark & \cmark & $-296.5$ & \ref{fig:cornerMBHB+SL} & \ref{tab:glitchGW1}\\
         8 &\textsc{D} & \cmark & \xmark & \cmark & $-48.8$ & \ref{fig:cornerXHM_SL_D} & \ref{tab:glitchGW3}\\
    \rowcolor{teal!20}     9 &\textsc{A1} & \cmark & \textsc{A1} & \cmark & $-46.8$ & \ref{fig:cornerMBHB+LPF} & \ref{tab:glitchGW2}\\
    \rowcolor{teal!20}     10 &\textsc{A2} & \cmark & \textsc{A2} & \cmark & $-43.9$ & \ref{fig:cornerMBHB+SL} & \ref{tab:glitchGW1}\\
    \rowcolor{teal!20}    11 &\textsc{D} & \cmark & \textsc{D} & \cmark & $-40.8$ & \ref{fig:cornerXHM_SL_D} & \ref{tab:glitchGW3}\\
         12 & \xmark & \cmark & \textsc{A1} & \cmark & $-75.0$ &  & \\
         13 & \xmark & \cmark & \textsc{A2} & \cmark & $-44.1$ &  & \\
         14 &\xmark & \cmark & \textsc{D} & \cmark & $-52.2$ &  & \\
    \end{tabular}
    \caption{
    Summary of our runs containing single glitches and the fiducial GW signal.
    Rows highlighted in teal 
    denote runs where the recovery signal model matches that of the injection.
    We first perform disjoint parameter estimation on our fiducial GW source and three glitch models (IDs 1-5). 
    We then generate signals from the superposition of a GW signal with single glitches and study them both ignoring (IDs 6-8)  and including (IDs 9-11) the glitch in the data model.
    Finally, we generate GW signals and perform parameter estimation on them including glitches in the data model (IDs 12-14).
    }
    \label{tab:summary-runs}
\end{table*}

\begin{table*}[ht]
    \centering
    \begin{tabular}
    {c | c | c | c | c | c | c | c | c | c | c | c}
         ID &\multicolumn{4}{c|}{Injection} & \multicolumn{4}{c|}{Recovery} & log-Evidence & Figure & Table\\
    \hline
         &\multicolumn{2}{c|}{Glitch 1} & \multicolumn{2}{c|}{Glitch 2} &
         \multicolumn{2}{c|}{Glitch 1} & \multicolumn{2}{c|}{Glitch 2}
         & & & \\
    \hline
        \rowcolor{teal!20} 15&$\textsc{D}(1,1)$ & $\textsc{D}(1,2)$ & \xmark & \xmark & $\textsc{D}(1,1)$ & $\textsc{D}(1,2)$ & \xmark & \xmark & $-16.1$ &  & \ref{tab:glitchDisp_param}\\
         16& &  & &  &  $\textsc{D}(1,1)$ & $\textsc{D}(1,3)$ & \xmark & \xmark & $-18.0$ &  & \\
        17& &  & &  &  $\textsc{D}(1,2)$ & $\textsc{D}(1,3)$ & \xmark & \xmark & $-20.1$ &  & \\
        18& & & &  &  $\textsc{D}(1,1)$ & \xmark & \xmark & \xmark & $-22.9$ &  & \\
        19& &  & &  & $\textsc{D}(1,2)$ & \xmark & \xmark & \xmark & $-23.9$ &  & \\
        20& &  & &  & $\textsc{D}(1,3)$ & \xmark & \xmark & \xmark & $-34.4$ &  & \\
        21& &  & &  & $\textsc{D}(1,1)$ & $\textsc{D}(1,2)$ & $\textsc{D}(1,3)$& \xmark & $-17.0$ &  & \\
        \hline
        \rowcolor{teal!20} 22&$\textsc{D}(1,1)$ & \xmark &\xmark & \xmark & $\textsc{D}(1,1)$ & \xmark & \xmark & \xmark & $-15.2$ & & \ref{tab:glitchDisp_param}\\
        23& & & & & $\textsc{D}(2,1)$ & \xmark & \xmark & \xmark & $-3650.20$ & & \\
        24& & & & & $\textsc{D}(3,1)$ & \xmark & \xmark & \xmark & $-3650.18$ &  & \\
        25& & & & & $\textsc{D}(1^\prime,1)$ & \xmark & \xmark & \xmark & $-224.1$ &  & \\
        26& & & & & $\textsc{D}(2^\prime,1)$ & \xmark & \xmark & \xmark & $-3628.6$ &  & \\
        27& & & & & $\textsc{D}(3^\prime,1)$ & \xmark & \xmark & \xmark & $-3640.8$ &  & \\
        \hline
        \rowcolor{teal!20} 28&$\textsc{D}(1,2)$ & $\textsc{D}(1,3)$ & $\textsc{D}(3,1)$ & $\textsc{D}(3,2)$ & $\textsc{D}(1,2)$ & $\textsc{D}(1,3)$ & $\textsc{D}(3,1)$ & $\textsc{D}(3,2)$ & $-34.8$ & \ref{fig:corner4D} & \ref{tab:glitchDisp_param}\\
        \hline
        \rowcolor{teal!20} 29&$\textsc{A2}(1,1)$ &\xmark  &$\textsc{A2}(2^\prime,2)$ & \xmark & $\textsc{A2}(1,1)$ &\xmark &  $\textsc{A2}(2^\prime,2)$ & \xmark & $-15.9$ & \ref{fig:cornerA2} & \ref{tab:glitchAcc_param}\\
        \hline
        \rowcolor{teal!20} 30&$\textsc{A1s}(1)$ & \xmark &\xmark & \xmark & $\textsc{A1s}(1)$ & \xmark & \xmark & \xmark & $-13.6$ & \ref{fig:cornerLPFlmh} & \ref{tab:glitchAcc_param}\\ 
        \hline
        \rowcolor{teal!20} 31&$\textsc{A1m}(1)$ & \xmark &\xmark & \xmark & $\textsc{A1m}(1)$ & \xmark & \xmark & \xmark & $-18.0$ & \ref{fig:cornerLPFlmh} & \ref{tab:glitchAcc_param}\\
        \hline
        \rowcolor{teal!20} 32&$\textsc{A1l}(1)$ & \xmark &\xmark & \xmark & $\textsc{A1l}(1)$ & \xmark & \xmark & \xmark & $-16.6$ & \ref{fig:cornerLPFlmh}& \ref{tab:glitchAcc_param}\\
    \end{tabular}
    \caption{
    Summary of a large set of injected glitches and associated recoveries. 
    Injected glitches are labeled by \textsc{X}(i,n), with \textsc{X}, $n$, and $i$ describing the glitch model, the injection point, and the shapelet order (when applicable), respectively.
    We explore the number of components and shapelet order (IDs 16-20), the number of glitches (ID 21), and the potential misidentification of the injection point (IDs 22-27).
    Additionally, we simulate data from glitches of increasing complexity (IDs 28, 29) and consider three representative glitches inspired by LPF data (IDs 30-32). These are a short duration and small amplitude glitch (\textsc{A1s}), a medium duration and amplitude glitch (\textsc{A1m}), and a long duration and large amplitude glitch (\textsc{A1l}).
  	Runs with same injected signals are grouped by horizontal lines. 
  }
    
    \label{tab:summary-runs-glitchonly}
\end{table*}

\section{Inference}
\label{sec:inference}

The initial search of a GW in noisy data is achieved through matched-filtering techniques~\cite{1933RSPTA.231..289N}
which provide initial guesses on the signal parameters.
If glitches are present, their preliminary detection and subtraction might not be sufficient to provide data that are sufficiently cleaned to accurately infer the parameters of the astrophysical source~\cite{2022PhRvD.105d2002B}. 
Previous studies presented a matching-pursuit algorithm for an automated and systematic glitch detection~\cite{1993ITSP...41.3397M} showing that, while the search grid on the damping parameter is too coarse to accurately obtain the best-fit glitch, it provides a reliable initial guess.
For practical purposes, here we assume that such guess has been identified from the data and can be used to inform our subsequent analyses.

We perform a joint parameter estimation, fitting simultaneously for GW signals and noise artifacts. 
We construct posteriors on parameters $\boldsymbol{\theta}$ 
\begin{equation}
    p(\boldsymbol{\theta}|d)\propto \mathcal{L}(d|\boldsymbol{\theta})\pi(\boldsymbol{\theta})
    \label{eq:PDF}
\end{equation}
through stochastic sampling of the likelihood $\mathcal{L}(d|\boldsymbol{\theta})$ under a prior $\pi(\boldsymbol{\theta})$.
We employ a coherent analysis on the three noise-orthogonal TDI channels $d = \{d_k; \textit{k} = M_A, M_E, M_T\}$ when considering displacement variables and  $d = \{d_k; \textit{k} = G_A, G_E, G_T\}$ when considering acceleration variables.
We use a Gaussian likelihood \cite{1994PhRvD..49.2658C}
\begin{equation}
    \text{ln}\mathcal{L}(d|\boldsymbol{\theta})=-\sum_{k}{\frac{({d}_{k}-{s}_{k}(\boldsymbol{\theta})|{d}_{k}-{s}_{k}(\boldsymbol{\theta}))_{k}}{2}+\text{const.}},
    \label{eq:LISA_likelihood}
\end{equation}
where $\Tilde{s}_{k}$ is the $k$-th TDI output frequency series associated to the injected signal $\Tilde{s}(f;\boldsymbol{\theta})$.
The output $\Tilde{s}_{k}$ represent either acceleration or fractional displacements depending on the chosen TDI variable set, thus containing acceleration glitches, 
displacement glitches,
GW transients,
or a combination of these (cf. Sec.~\ref{sec:model}).
The noise-weighted inner product is defined as 
\begin{equation}
    (a \mid b)_{k}=4\Re\int_{f_\mathrm{min}}^{f_\mathrm{max}}{\frac{\tilde{a}^{*}(f)\tilde{b}(f)}{S_{k}(f)}df},
    \label{eq:inner_product}
\end{equation}
where $\Re$ denotes the real part, $\Tilde{a}(f)$ is the Fourier transform of the time series $a(t)$, and $S_{k}(f)$ is the one-sided noise spectral density of the $k$-th TDI channel. 
We use the match between two signals
\begin{equation} \label{eq:match}
M(a,b) = \frac{(a\mid b)}{(a\mid a)^{1/2}(b\mid b)^{1/2}}
\end{equation}
to optimize the onset time of the injected glitches as discussed in Sec.~\ref{subsec:gw-model}.
Model selection is performed using log-Bayes factors %$\log_{10}\mathcal{B}_i^j$ 
\begin{equation}
\label{eq:BF}
    \log_{10}\mathcal{B}_i^j = \log_{10} \mathcal{Z}_i - \log_{10} \mathcal{Z}_j, 
\end{equation}
where  $i$ and $j$ are labels identifying the competing models, and %, or log-marginal-likelihood
\begin{equation}
    \mathcal{Z}(d) = \int d \boldsymbol{\theta} \mathcal{L}(d\mid \boldsymbol{\theta}) \pi(\boldsymbol{\theta})
\end{equation}
is the evidence of each parameter estimation. 

We consider a LISA mission lifetime of $T_{\textsc{LISA}}=4$ years, roughly equivalent to a calendar observation time of 4.5 years with an effective duty cycle of 82\%. 
Our frequency resolution is therefore $\Delta f\approx1/T_{\textsc{LISA}}=1.7\times10^{-8}\,\rm{Hz}$.
We set  $f_{\mathrm{min}}=0.1\,\rm{mHz}$ and $f_{\mathrm{max}}= 4\,\rm{mHz}$, which is well above the fiducial GW and the maximum frequencies of all glitch signals.
We use a semi-analytical noise spectral density model $S_{k}(f)$~\cite{SciRD} describing the superposition of LISA stationary instrumental noise and astrophysical confusion noise from unresolved Galactic binaries \cite{2017PhRvD..95j3012B}.
In order to reduce the computational cost, we evaluate inner products from Eq.~\eqref{eq:inner_product} using a Clenshaw-Curtis integration algorithm~\cite{CLENSHAW1960}, see e.g~Ref.~\cite{2021PhRvD.104d4065B} for a summary of its application to LISA data.
    
Parameter estimation is performed with the \textsc{Balrog} code, which is designed to work with different stochastic samplers. 
In particular, in this paper we use the nested sampling algorithm~\cite{2004AIPC..735..395S} as implemented in \textsc{Nessai}~\cite{2021PhRvD.103j3006W}.  
We choose uniform priors on each parameter over either its entire definition domain or a range that is sufficiently large to enclose the entire posterior.

\section{\label{sec:results} Results}

We perform two sets of parameter-estimation runs: 
\begin{enumerate}
\item[(i)] Joint inference runs on both GW signal and glitches (Sec.~\ref{subsec:gw+glitch}), listed with IDs 1 to 14 in Table~\ref{tab:summary-runs};
\item[(ii)] Inference runs where we inject and recover glitches without GW signal (Sec.~\ref{subsec:glitchonly}), listed with IDs 15 to 32 in Table~\ref{tab:summary-runs-glitchonly}.
\end{enumerate}

\begin{table*}[ht]
    \centering    
    \renewcommand\arraystretch{1.8}
    \resizebox{\textwidth}{!}{
    \begin{tabular}
    {c | c| c| c| c| c| c| c| c| c| c| c| c| c| c | c}
    & \multicolumn{11}{c}{MBHB} &
    \multicolumn{4}{|c}{\textsc{Model A1}}
    \\
    \hline
        ID &
        $m_{1z}~[10^7 M_\odot]$ &
        $m_{2z}~[10^7 M_\odot]$ &
        $t_m~[\rm{h}]$ &
        $\chi_1$ &
        $\chi_2$ &
        $d_L~[\rm{Gpc}]$ &
        $\iota~[\rm{rad}]$ &
        $\beta~[\rm{rad}]$ &
        $\lambda~[\rm{rad}]$ &
        $\phi~[\rm{rad}]$ &
        $\psi~[\rm{rad}]$ &
        $A~[\rm{pm/s}]$& 
        $\beta_1~[\rm{s}]$&
        $\beta_2~[\rm{s}]$& 
        $\tau~[\rm{h}]$
        \\
    \hline
    &
        \RunNineInjBHBOneRedshiftedMassOneZero &
        \RunNineInjBHBOneRedshiftedMassTwoZero &
        \RunNineInjBHBOneMergerTimeOrInitialOrbitalFrequencyZero &
        \RunNineInjBHBOneDimensionlessSpinOneZero &
        \RunNineInjBHBOneDimensionlessSpinTwoZero &
        \RunNineInjBHBOneLuminosityDistanceZero &
        \RunNineInjBHBOneInclinationZero &
        \RunNineInjBHBOneEclipticLatitudeZero &
        \RunNineInjBHBOneEclipticLongitudeZero &
        \RunNineInjBHBOneInitialOrbitalPhaseZero &
        \RunNineInjBHBOnePolarizationZero &
        \RunNineInjLPFOneDeltavZero &
        \RunNineInjLPFOneTauRiseZero &
        \RunNineInjLPFOneTauFallZero &
        \RunNineInjLPFOneInitialTimeZero
        \\
    \rowcolor{teal!20}
    9 &
        \RunNineBHBOneRedshiftedMassOneZero &
        \RunNineBHBOneRedshiftedMassTwoZero &
        \RunNineBHBOneMergerTimeOrInitialOrbitalFrequencyZero &
        \RunNineBHBOneDimensionlessSpinOneZero &
        \RunNineBHBOneDimensionlessSpinTwoZero &
        \RunNineBHBOneLuminosityDistanceZero &
        \RunNineBHBOneInclinationZero &
        \RunNineBHBOneEclipticLatitudeZero &
        \RunNineBHBOneEclipticLongitudeZero &
        \RunNineBHBOneInitialOrbitalPhaseZero &
        \RunNineBHBOnePolarizationZero &
        \RunNineLPFOneDeltavZero &
        \RunNineLPFOneTauRiseZero &
        \RunNineLPFOneTauFallZero &
        \RunNineLPFOneInitialTimeZero
        \\
    \rowcolor{teal!20}
    6 &
        \RunSixBHBOneRedshiftedMassOneZero &
        \RunSixBHBOneRedshiftedMassTwoZero &
        \RunSixBHBOneMergerTimeOrInitialOrbitalFrequencyZero &
        \RunSixBHBOneDimensionlessSpinOneZero &
        \RunSixBHBOneDimensionlessSpinTwoZero &
        \RunSixBHBOneLuminosityDistanceZero &
        \RunSixBHBOneInclinationZero &
        \RunSixBHBOneEclipticLatitudeZero &
        \RunSixBHBOneEclipticLongitudeZero &
        \RunSixBHBOneInitialOrbitalPhaseZero &
        \RunSixBHBOnePolarizationZero &
        \xmark &
        \xmark &
        \xmark &
        \xmark
        \\
        &
        \RunSixInjBHBOneRedshiftedMassOneZero &
        \RunSixInjBHBOneRedshiftedMassTwoZero &
        \RunSixInjBHBOneMergerTimeOrInitialOrbitalFrequencyZero &
        \RunSixInjBHBOneDimensionlessSpinOneZero &
        \RunSixInjBHBOneDimensionlessSpinTwoZero &
        \RunSixInjBHBOneLuminosityDistanceZero &
        \RunNineInjBHBOneInclinationZero &
        \RunNineInjBHBOneEclipticLatitudeZero &
        \RunSixInjBHBOneEclipticLongitudeZero &
        \RunSixInjBHBOneInitialOrbitalPhaseZero &
        \RunSixInjBHBOnePolarizationZero &
        \xmark &
        \xmark &
        \xmark &
        \xmark
        \\
    \rowcolor{gray!20} 1 &
        \RunOneBHBOneRedshiftedMassOneZero &
        \RunOneBHBOneRedshiftedMassTwoZero &
        \RunOneBHBOneMergerTimeOrInitialOrbitalFrequencyZero &
        \RunOneBHBOneDimensionlessSpinOneZero &
        \RunOneBHBOneDimensionlessSpinTwoZero &
        \RunOneBHBOneLuminosityDistanceZero &
        \RunOneBHBOneInclinationZero &
        \RunTenBHBOneEclipticLatitudeZero &
        \RunOneBHBOneEclipticLongitudeZero &
        \RunOneBHBOneInitialOrbitalPhaseZero &
        \RunOneBHBOnePolarizationZero &
        \xmark &
        \xmark &
        \xmark &
        \xmark
    \end{tabular}}
    \caption{
    Parameter estimation results for a GW signal contaminated by a \textsc{Model A1} glitch. 
    The injected parameters are listed in the white rows.
    Medians and 90\% credible intervals for the recovered posteriors are listed in the two rows highlighted in teal. 
    While accounting for the presence of a glitch (ID 9) allows for joint unbiased reconstruction of all parameters, ignoring its potential occurrence (ID 6) yields large systematic biases.
    Ignoring the presence of a glitch is disfavored with $\log_{10}\mathcal{B}^{6}_{9}= -14491$.
    Joint posterior distributions for both these runs are shown in Fig.~\ref{fig:cornerMBHB+LPF}.
    For comparison, the bottom row shows our reference run where we only inject the GW source (ID 1). The subset of parameters common across runs 1 and 9 does not show appreciable differences. 
    }\label{tab:glitchGW2}
    
    \bigskip
    
    \renewcommand\arraystretch{1.2}
    \resizebox{\textwidth}{!}{%
    \begin{tabular}
    {c | c| c| c| c| c| c| c| c| c| c| c| c| c| c}
    & \multicolumn{11}{c}{MBHB} &
    \multicolumn{3}{|c}{\textsc{Model A2}}
    \\
    \hline
    ID &
        $m_{1z}~[10^7 M_\odot]$ &
        $m_{2z}~[10^7 M_\odot]$ &
        $t_m~[\rm{h}]$ &
        $\chi_1$ &
        $\chi_2$ &
        $d_L~[\rm{Gpc}]$ &
        $\iota~[\rm{rad}]$ &
        $\beta~[\rm{rad}]$ &
        $\lambda~[\rm{rad}]$ &
        $\phi~[\rm{rad}]$ &
        $\psi~[\rm{rad}]$ &
        $A~[\rm{pm/s}]$& 
        $\beta~[\rm{s}]$& 
        $\tau~[\rm{h}]$
        \\
    \hline
    &
        \RunTenInjBHBOneRedshiftedMassOneZero &
        \RunTenInjBHBOneRedshiftedMassTwoZero &
        \RunTenInjBHBOneMergerTimeOrInitialOrbitalFrequencyZero &
        \RunTenInjBHBOneDimensionlessSpinOneZero &
        \RunTenInjBHBOneDimensionlessSpinTwoZero &
        \RunTenInjBHBOneLuminosityDistanceZero &
        \RunNineInjBHBOneInclinationZero &
        \RunNineInjBHBOneEclipticLatitudeZero &
        \RunTenInjBHBOneEclipticLongitudeZero &
        \RunTenInjBHBOneInitialOrbitalPhaseZero &
        \RunTenInjBHBOnePolarizationZero &
        \RunTenInjSLOneAmplitudeDisplacementZero &
        \RunTenInjSLOneBetaZero &
        \RunTenInjSLOneInitialTimeZero
        \\
    \rowcolor{teal!20}
    10 &
        \RunTenBHBOneRedshiftedMassOneZero &
        \RunTenBHBOneRedshiftedMassTwoZero &
        \RunTenBHBOneMergerTimeOrInitialOrbitalFrequencyZero &
        \RunTenBHBOneDimensionlessSpinOneZero &
        \RunTenBHBOneDimensionlessSpinTwoZero &
        \RunTenBHBOneLuminosityDistanceZero &
        \RunTenBHBOneInclinationZero &
        \RunTenBHBOneEclipticLatitudeZero &
        \RunTenBHBOneEclipticLongitudeZero &
        \RunTenBHBOneInitialOrbitalPhaseZero &
        \RunTenBHBOnePolarizationZero &
        \RunTenSLOneAmplitudeDisplacementZero &
        \RunTenSLOneBetaZero &
        \RunTenSLOneInitialTimeZero
        \\
    \rowcolor{teal!20}
    7 &
        \RunSevenBHBOneRedshiftedMassOneZero &
        \RunSevenBHBOneRedshiftedMassTwoZero &
        \RunSevenBHBOneMergerTimeOrInitialOrbitalFrequencyZero &
        \RunSevenBHBOneDimensionlessSpinOneZero &
        \RunSevenBHBOneDimensionlessSpinTwoZero &
        \RunSevenBHBOneLuminosityDistanceZero &
        \RunSevenBHBOneInclinationZero &
        \RunSevenBHBOneEclipticLatitudeZero &
        \RunSevenBHBOneEclipticLongitudeZero &
        \RunSevenBHBOneInitialOrbitalPhaseZero &
        \RunSevenBHBOnePolarizationZero &
        \xmark &
        \xmark &
        \xmark
    \\
            &
        \RunOneInjBHBOneRedshiftedMassOneZero &
        \RunOneInjBHBOneRedshiftedMassTwoZero &
        \RunOneInjBHBOneMergerTimeOrInitialOrbitalFrequencyZero &
        \RunOneInjBHBOneDimensionlessSpinOneZero &
        \RunOneInjBHBOneDimensionlessSpinTwoZero &
        \RunOneInjBHBOneLuminosityDistanceZero &
        \RunOneInjBHBOneInclinationZero &
        \RunNineInjBHBOneEclipticLatitudeZero &
        \RunOneInjBHBOneEclipticLongitudeZero &
        \RunOneInjBHBOneInitialOrbitalPhaseZero &
        \RunOneInjBHBOnePolarizationZero &
        \xmark &
        \xmark &
        \xmark 
        \\
    \rowcolor{gray!20} 1 &
        \RunOneBHBOneRedshiftedMassOneZero &
        \RunOneBHBOneRedshiftedMassTwoZero &
        \RunOneBHBOneMergerTimeOrInitialOrbitalFrequencyZero &
        \RunOneBHBOneDimensionlessSpinOneZero &
        \RunOneBHBOneDimensionlessSpinTwoZero &
        \RunOneBHBOneLuminosityDistanceZero &
        \RunOneBHBOneInclinationZero &
        \RunTenBHBOneEclipticLatitudeZero &
        \RunOneBHBOneEclipticLongitudeZero &
        \RunOneBHBOneInitialOrbitalPhaseZero &
        \RunOneBHBOnePolarizationZero &
        \xmark &
        \xmark &
        \xmark 
    \end{tabular}}
    \caption{
    Parameter estimation results for a GW signal contaminated by a \textsc{Model A2} glitch. 
    Results are organized as in Table~\ref{tab:glitchGW2}. 
    This glitch, if present in data and ignored upon inference, introduces milder biases when compared to run with ID 6: 
    this is due to its shorter duration resulting in a smaller match with the GW waveform. 
    Joint posterior distributions for both runs are shown in Fig.~\ref{fig:cornerMBHB+SL}.
}\label{tab:glitchGW1}
        
        \bigskip
        
    \renewcommand\arraystretch{1.8}
    \resizebox{\textwidth}{!}{%
    \begin{tabular}
    {c | c| c| c| c| c| c| c| c| c| c| c| c| c| c}
    & \multicolumn{11}{c}{MBHB} &
    \multicolumn{3}{|c}{\textsc{Model D}}
    \\
    \hline
        ID &
        $m_{1z}\!~[10^7 M_\odot]$ &
        $m_{2z}\!~[10^7 M_\odot]$ &
        $t_m~[\rm{h}]$ &
        $\chi_1$ &
        $\chi_2$ &
        $d_L~[\rm{Gpc}]$ &
        $\iota~[\rm{rad}]$ &
        $\beta~[\rm{rad}]$ &
        $\lambda~[\rm{rad}]$ &
        $\phi~[\rm{rad}]$ &
        $\psi~[\rm{rad}]$ &
        $D~[\rm{pm\!\cdot\!s}]$& 
        $\beta~[\rm{s}]$&
        $\tau~[\rm{h}]$
        \\
    \hline
    &
        \RunElevenInjBHBOneRedshiftedMassOneZero &
        \RunElevenInjBHBOneRedshiftedMassTwoZero &
        \RunElevenInjBHBOneMergerTimeOrInitialOrbitalFrequencyZero &
        \RunElevenInjBHBOneDimensionlessSpinOneZero &
        \RunElevenInjBHBOneDimensionlessSpinTwoZero &
        \RunElevenInjBHBOneLuminosityDistanceZero &
        \RunElevenInjBHBOneInclinationZero &
        \RunTenInjBHBOneEclipticLatitudeZero &
        \RunElevenInjBHBOneEclipticLongitudeZero &
        \RunElevenInjBHBOneInitialOrbitalPhaseZero &
        \RunElevenInjBHBOnePolarizationZero &
        \RunElevenInjSLOneAmplitudeDisplacementZero &
        \RunElevenInjSLOneBetaZero &
        \RunElevenInjSLOneInitialTimeZero
        \\
    \rowcolor{teal!20}
    11 &
        \RunElevenBHBOneRedshiftedMassOneZero &
        \RunElevenBHBOneRedshiftedMassTwoZero &
        \RunElevenBHBOneMergerTimeOrInitialOrbitalFrequencyZero &
        \RunElevenBHBOneDimensionlessSpinOneZero &
        \RunElevenBHBOneDimensionlessSpinTwoZero &
        \RunElevenBHBOneLuminosityDistanceZero &
        \RunElevenBHBOneInclinationZero &
        \RunElevenBHBOneEclipticLatitudeZero &
        \RunElevenBHBOneEclipticLongitudeZero &
        \RunElevenBHBOneInitialOrbitalPhaseZero &
        \RunElevenBHBOnePolarizationZero &
        \RunElevenSLOneAmplitudeDisplacementZero &
        \RunElevenSLOneBetaZero &
        \RunElevenSLOneInitialTimeZero
        \\
    \rowcolor{teal!20}
    8 &
        \RunEightBHBOneRedshiftedMassOneZero &
        \RunEightBHBOneRedshiftedMassTwoZero &
        \RunEightBHBOneMergerTimeOrInitialOrbitalFrequencyZero &
        \RunEightBHBOneDimensionlessSpinOneZero &
        \RunEightBHBOneDimensionlessSpinTwoZero &
        \RunEightBHBOneLuminosityDistanceZero &
        \RunEightBHBOneInclinationZero &
        \RunEightBHBOneEclipticLatitudeZero &
        \RunEightBHBOneEclipticLongitudeZero &
        \RunEightBHBOneInitialOrbitalPhaseZero &
        \RunEightBHBOnePolarizationZero &
        \xmark &
        \xmark &
        \xmark
    \\
     &
        \RunTwoInjBHBOneRedshiftedMassOneZero &
        \RunTwoInjBHBOneRedshiftedMassTwoZero &
        \RunTwoInjBHBOneMergerTimeOrInitialOrbitalFrequencyZero &
        \RunTwoInjBHBOneDimensionlessSpinOneZero &
        \RunTwoInjBHBOneDimensionlessSpinTwoZero &
        \RunTwoInjBHBOneLuminosityDistanceZero &
        \RunTenInjBHBOneInclinationZero &
        \RunTenInjBHBOneEclipticLatitudeZero &
        \RunTwoInjBHBOneEclipticLongitudeZero &
        \RunTwoInjBHBOneInitialOrbitalPhaseZero &
        \RunTwoInjBHBOnePolarizationZero &
        \xmark &
        \xmark &
        \xmark
        \\
    \rowcolor{gray!20}
    2 &
        \RunTwoBHBOneRedshiftedMassOneZero &
        \RunTwoBHBOneRedshiftedMassTwoZero &
        \RunTwoBHBOneMergerTimeOrInitialOrbitalFrequencyZero &
        \RunTwoBHBOneDimensionlessSpinOneZero &
        \RunTwoBHBOneDimensionlessSpinTwoZero &
        \RunTwoBHBOneLuminosityDistanceZero &
        \RunTwoBHBOneInclinationZero &
        \RunTwoBHBOneEclipticLatitudeZero &
        \RunTwoBHBOneEclipticLongitudeZero &
        \RunTwoBHBOneInitialOrbitalPhaseZero &
        \RunTwoBHBOnePolarizationZero &
        \xmark &
        \xmark &
        \xmark
    \end{tabular}}
    \caption{
    Parameter estimation results for a GW signal contaminated by a \textsc{Model D} glitch. 
    Results are organized as in Table~\ref{tab:glitchGW2}. 
    This glitch, if present in data and ignored upon inference, introduces negligible biases when compared to runs with IDs 6 and 7. 
    This is due to its very short duration, which superimposes with the GW signal only for a few seconds yielding a low match. 
    Joint posterior distributions for both runs are shown in Fig.~\ref{fig:cornerXHM_SL_D}.
    }\label{tab:glitchGW3}
    \end{table*}

\subsection[GW + Glitch]{\label{subsec:gw+glitch} Joint inference with glitches and GWs}

If a preliminary search fails to identify and remove a glitch from the data, it is important to assess its impact on the parameters of the overlapping GW source. We thus tackle the following cases for each of the three signals illustrated in Fig.~\ref{fig:Waveform}:
\begin{itemize}
\item Parameter estimation in the absence of glitch in the data (``reference'' runs, with IDs 1 and 2);
\item Parameter estimation ignoring a glitch when present in the data (``glitch-ignorant'' runs, with IDs 6-8);
\item Parameter estimation including in the signal model a glitch that is present in the data (``glitch-complete'' runs, with IDs 9-11). 
\end{itemize}
Bayesian evidence for each run is listed Tab.~\ref{tab:summary-runs}.
We report $\log_{10}\mathcal{B}_{9}^{6}$, $\log_{10}\mathcal{B}_{10}^{7}$, and $\log_{10}\mathcal{B}_{11}^{8}$ much greater than 2, indicating a ``decisive" evidence~\cite{10.2307/2291091} in favor of a glitch being present in the data.

\begin{figure*}[p]
    \centering
    \includegraphics[width=\textwidth, keepaspectratio]{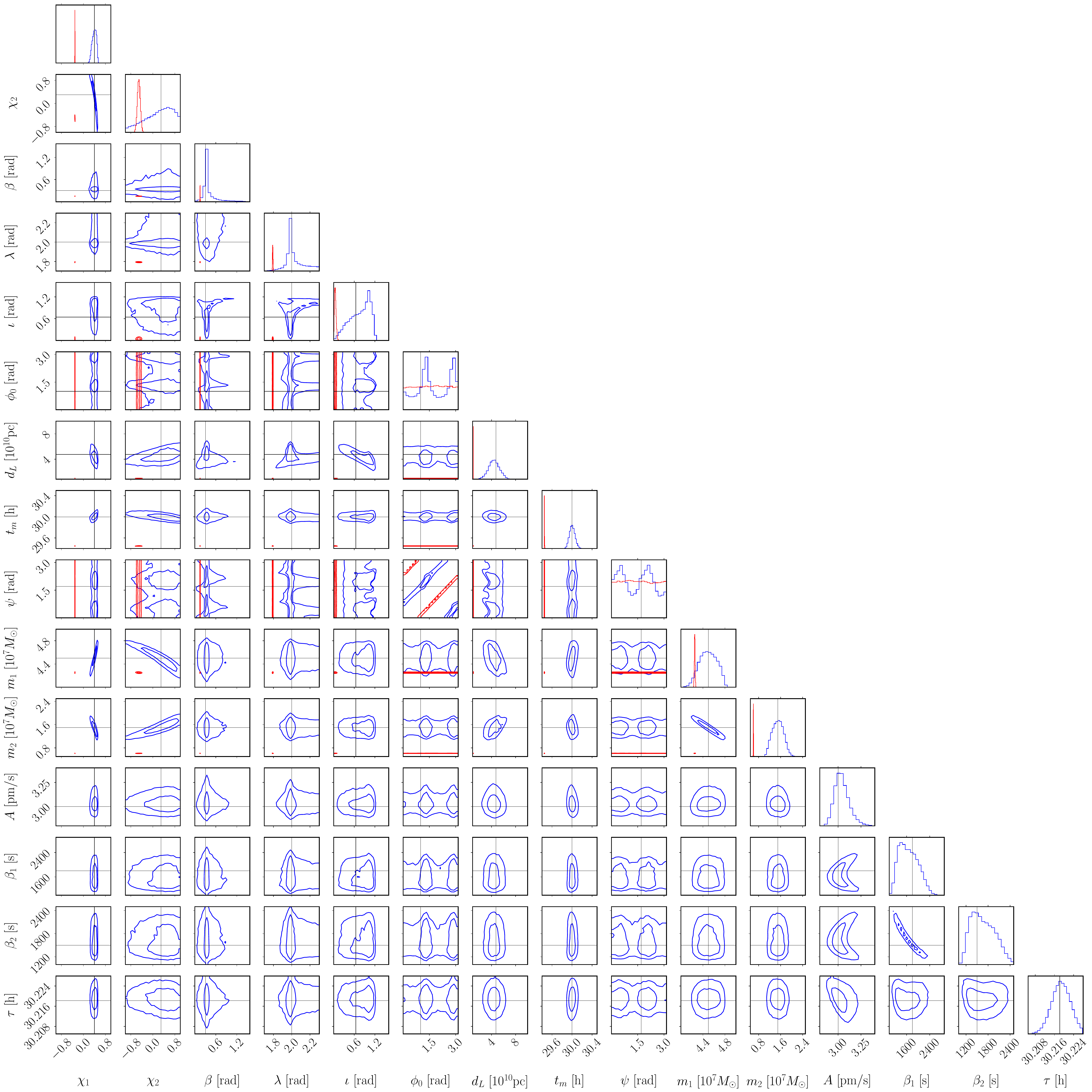}
    \caption{
   Posterior distribution in blue (red) corresponding to run ID 9 (6) where a  \textsc{Model A1} glitch is (not) included in the recovery process. 
   Contours indicate the 50\% and 90\% credible regions; solid black lines indicate the injected values as listed in Table~\ref{tab:glitchGW2}.
   When the glitch is included in the inference, each model injected parameter is recovered within the 90\% one-dimensional credible region. 
   We do not report notable correlations between glitch and GW parameters. 
   If the glitch is excluded, all MBHB parameters except the initial phase $\phi_0$ and the polarization angle $\psi$ are systematically biased. In particular, the posterior on the luminosity distance $d_L$ rails heavily against the prior lower bound.}
    \label{fig:cornerMBHB+LPF}
\end{figure*}

\begin{figure*}[p]
    \centering
    \includegraphics[width=\textwidth, keepaspectratio]{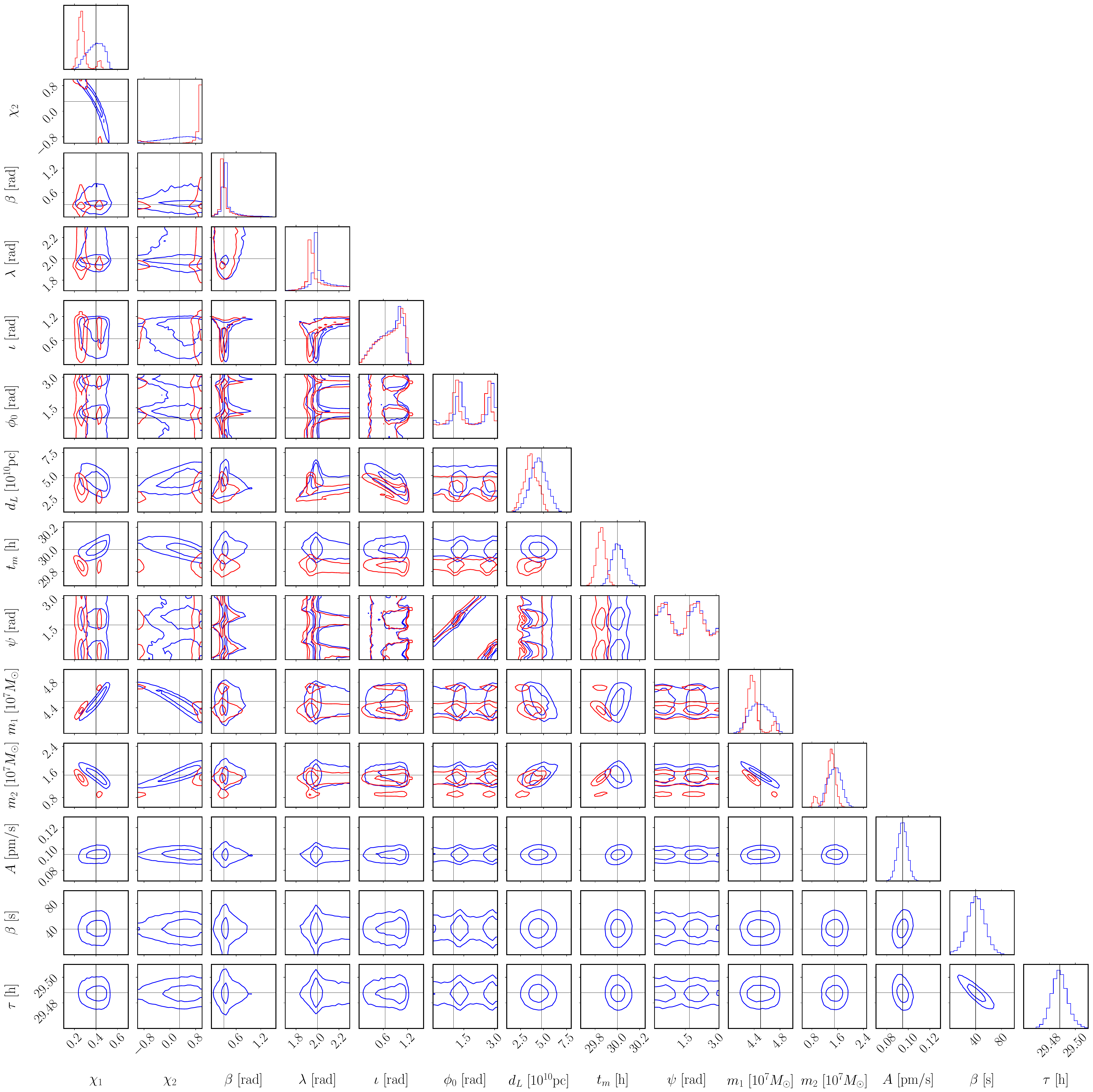}
    \caption{
Posterior distribution in blue (red) corresponding to run ID 10 (7) where a  \textsc{Model A2} glitch is (not) included in the recovery process. 
   Contours indicate the 50\% and 90\% credible regions; solid black lines indicate the injected values as listed in Table~\ref{tab:glitchGW1}.
 When the glitch is ignored, the MBHB parameters are somewhat biased; see  in particular the black-hole masses and spins.
    When the glitch is included in the recovery process, all model parameters are recovered within their 90\% one-dimensional credible regions. 
We do not report notable correlations between glitch and GW parameters.
    }
    \label{fig:cornerMBHB+SL}
\end{figure*}

\begin{figure*}[p]
    \centering
    \includegraphics[width=\textwidth, keepaspectratio]{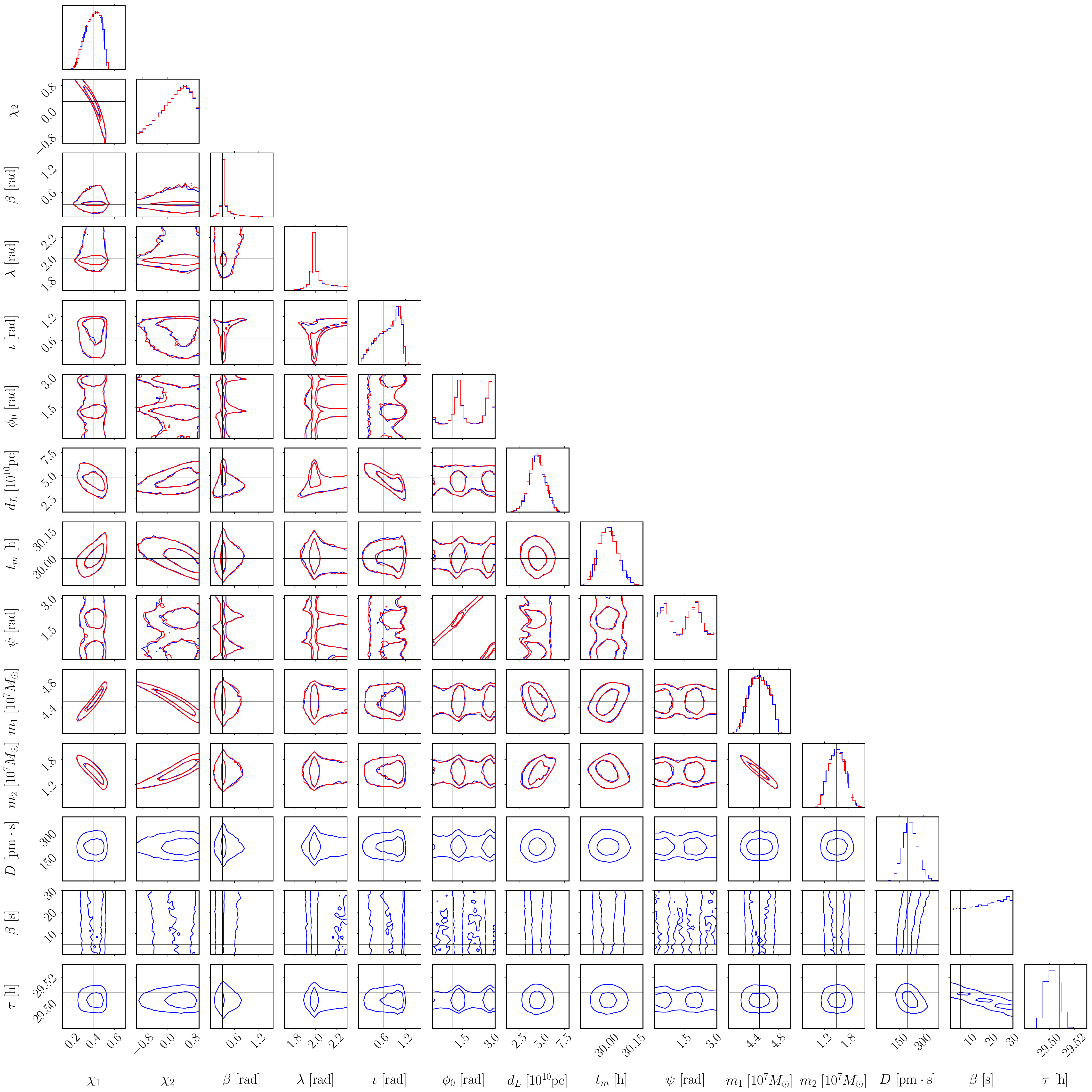}
    \caption{
    Posterior distribution in blue (red) corresponding to run ID 11 (8) where a \textsc{Model D} glitch is (not) included in the recovery process. 
   Contours indicate the 50\% and 90\% credible regions; solid black lines indicate the injected values as listed in Table~\ref{tab:glitchGW3}.
    When the glitch is ignored, the MBHB parameters are very mildly biased.
    In both cases, all model parameters are recovered within their 90\% one-dimensional credible regions. 
    }
    \label{fig:cornerXHM_SL_D}
\end{figure*}

Summaries are provided in Tables~\ref{tab:glitchGW2}, \ref{tab:glitchGW1}, and \ref{tab:glitchGW3}.
We find no appreciable differences in the posterior distribution of the GW-source parameters when comparing reference runs and glitch-complete runs, which is encouraging for LISA science. Individual parameters are well reconstructed, which is expected given the brightness of the source (SNR $\simeq 187$).
In particular, the MBHB component masses, the primary aligned spin component, and time to merger are measured with an accuracy of $\Delta m_i /m_i \approx 8-40\%$, $\Delta \chi_1 \approx 0.2$, and $\Delta t_m \approx 600\,\rm{s}$ (where we quote the 90\% credible interval of the marginal posterior distributions).
Figures~\ref{fig:cornerMBHB+LPF}, \ref{fig:cornerMBHB+SL}, and \ref{fig:cornerXHM_SL_D} show the posterior distribution for the fiducial MBHB of each glitch-complete run. Similarly, we do not report any appreciable difference with either fractional displacement or acceleration TDIs to model the same GW signal (see runs 1 and 2).

On the contrary, glitch-ignorant runs point to a different conclusion.
The resulting posterior depends on the chosen duration and amplitude of each transient (see runs 6, 7, and 8).
We find a long-duration, small-amplitude \textsc{Model A1} glitch massively contaminates the reconstruction of the GW parameters, to a point that the signal cannot be recovered at all.
This is shown in Fig.~\ref{fig:cornerMBHB+LPF}, where the glitch-ignorant distribution (red) shows evident issues in the underlying stochastic-sampling procedure.  This has to be contrasted with the regularity of the glitch-complete posterior distribution (blue), where instead the 
parameters of both GW signal and noise transient are successfully recovered. 
In particular, when the glitch is ignored we find that the posterior on the luminosity distance rails heavily against the lower bound of its prior, thus making the GW source reconstruction highly biased, even in a parameter space that largely encloses the posterior of the glitch-complete run.

As shown in Fig.~\ref{fig:cornerMBHB+SL}, a \textsc{Model A2} glitch 
with moderate duration and amplitude induces milder biases. Although the posterior support is far from the prior boundaries, the injected values lie outside the 99\% credible interval for both mass and spin parameters.
For the merger time,  the true value lies on the $97\%$ confidence interval of the corresponding marginalized posterior distribution.
The injected values of polarization, initial phase, inclination, and source position are within their one-dimensional 90\% confidence interval.

Equivalent runs for a \textsc{Model D} glitch are shown in Fig.~\ref{fig:cornerXHM_SL_D}. This is a noise transient that overlaps with the GW signal only for a small fraction of a cycle. As expected, we find such a glitch does not significantly impact the measurement of the GW parameters. 

Finally, we note that our glitch-complete runs do not exhibit significant cross-correlations between the glitch and GW parameters, thus effectively decoupling the inference on the two signals.

    \begin{table*}[t]
    \centering
    {
    \renewcommand\arraystretch{1.2}
    \rowcolors{2}{white}{teal!20}
    \resizebox{\textwidth}{!}{%
    \begin{tabular}
    {c|c|c|c|c|c|c|c|c|c|c|c}
    \rowcolor{white}& 
    \multicolumn{4}{c}{\textsc{Model A1}} &
    \multicolumn{6}{|c}{\textsc{Model A2}} \\
    \hline
    \rowcolor{white}
        ID & $A~[\rm{pm}/\rm{s}]$ & $\beta_1~[\rm{s}]$ & $\beta_2~[\rm{s}]$ & $\tau~[\rm{h}]$ & $A_0~[\rm{pm}/\rm{s}]$ & $\beta_0~[\rm{s}]$ & $\tau_0~[\rm{h}]$ & $A_1~[\rm{pm}/\rm{s}]$ & $\beta_1~[\rm{s}]$ & $\tau_1~[\rm{h}]$ \\
    \hline
    \rowcolor{white}    
     & \xmark & \xmark & \xmark & \xmark 
     & \RunTwentyNineInjSLOneAmplitudeDisplacementZero 
     & \RunTwentyNineInjSLOneBetaZero 
     & \RunTwentyNineInjSLOneInitialTimeZero 
     & \RunTwentyNineInjSLTwoAmplitudeDisplacementZero 
     & \RunTwentyNineInjSLTwoBetaZero
     & \RunTwentyNineInjSLTwoInitialTimeZero \\
    29 & \xmark & \xmark & \xmark & \xmark 
    & \RunTwentyNineSLOneAmplitudeDisplacementZero 
    & \RunTwentyNineSLOneBetaZero 
    & \RunTwentyNineSLOneInitialTimeZero
    & \RunTwentyNineSLTwoAmplitudeDisplacementZero 
    & \RunTwentyNineSLTwoBetaZero
    & \RunTwentyNineSLTwoInitialTimeZero \\
    \hline
         &
        \RunThirtyInjLPFOneDeltavZero &
        \RunThirtyInjLPFOneTauFallZero & 
        \RunThirtyInjLPFOneTauRiseZero & \RunThirtyInjLPFOneInitialTimeZero &
        \xmark & \xmark & \xmark & \xmark & \xmark & \xmark \\
        30 &
        \RunThirtyLPFOneDeltavZero &
        \RunThirtyLPFOneTauFallZero & 
        \RunThirtyLPFOneTauRiseZero & \RunThirtyLPFOneInitialTimeZero &
        \xmark & \xmark & \xmark & \xmark & \xmark & \xmark\\
    \hline
         &
        \RunThirtyOneInjLPFOneDeltavZero & \RunThirtyOneInjLPFOneTauFallZero & \RunThirtyOneInjLPFOneTauRiseZero & \RunThirtyOneInjLPFOneInitialTimeZero &
        \xmark & \xmark & \xmark & \xmark & \xmark & \xmark\\
        31 &
        \RunThirtyOneLPFOneDeltavZero & \RunThirtyOneLPFOneTauFallZero & \RunThirtyOneLPFOneTauRiseZero & \RunThirtyOneLPFOneInitialTimeZero &
        \xmark & \xmark & \xmark & \xmark & \xmark & \xmark\\
    \hline
        &
        \RunThirtyTwoInjLPFOneDeltavZero & \RunThirtyTwoInjLPFOneTauFallZero & \RunThirtyTwoInjLPFOneTauRiseZero & \RunThirtyTwoInjLPFOneInitialTimeZero &
        \xmark & \xmark & \xmark  & \xmark & \xmark & \xmark\\
        32 &
        \RunThirtyTwoLPFOneDeltavZero & \RunThirtyTwoLPFOneTauFallZero & \RunThirtyTwoLPFOneTauRiseZero & \RunThirtyTwoLPFOneInitialTimeZero &
        \xmark & \xmark & \xmark & \xmark & \xmark & \xmark\\
    \end{tabular}
    }
    }
    \caption{
    Parameter-estimation results on \textsc{Model A1} (ID 30-32) and \textsc{Model A2} (ID 29) glitches.  In particular, the former corresponds to glitches inspired by LPF observations, with varying duration and amplitudes.
    White rows show the injected values and teal rows show the recovered median and 90\% confidence interval.  
    The posterior distribution for these runs is provided in Figs.~\ref{fig:cornerLPFlmh} and \ref{fig:cornerA2}.
    }\label{tab:glitchAcc_param}
        
        \bigskip
        
    \rowcolors{2}{teal!20}{white}
    \renewcommand\arraystretch{1.8}
    \resizebox{\textwidth}{!}{%
    \begin{tabular}{c|c|c|c|c|c|c|c|c|c|c|c|c}
    \rowcolor{white} & 
    \multicolumn{6}{c}{Glitch 1} &
    \multicolumn{6}{|c}{Glitch 2} \\
    \hline
    \rowcolor{white} &
\multicolumn{3}{|c}{Component 1} &
\multicolumn{3}{|c}{Component 2} &
\multicolumn{3}{|c}{Component 1} &
\multicolumn{3}{|c}{Component 2} \\
    \hline
    \rowcolor{white}
        ID & $D_0~[\rm{pm}\cdot s]$ & $\beta_0~[\rm{s}]$ & $\tau_0~[\rm{h}]$ & $D_1~[\rm{pm}\cdot s]$ & $\beta_1~[\rm{s}]$ & $\tau_1~[\rm{h}]$ &$D_2~[\rm{pm}\cdot s]$ & $\beta_2~[\rm{s}]$ & $\tau_2~[\rm{h}]$ & $D_3~[\rm{pm}\cdot s]$ & $\beta_3~[\rm{s}]$ & $\tau_3~[\rm{h}]$ \\
    \hline
         & 
        \RunTwentyTwoInjSLOneAmplitudeDisplacementZero & \RunTwentyTwoInjSLOneBetaZero & \RunTwentyTwoInjSLOneInitialTimeZero &
        \xmark & \xmark & \xmark & \xmark & \xmark & \xmark & \xmark & \xmark & \xmark\\
        22 & 
        \RunTwentyTwoSLOneAmplitudeDisplacementZero & \RunTwentyTwoSLOneBetaZero & \RunTwentyTwoSLOneInitialTimeZero &
        \xmark & \xmark & \xmark & \xmark & \xmark & \xmark & \xmark & \xmark & \xmark\\
    \hline
         & 
        \RunFifteenInjSLOneAmplitudeDisplacementZero & \RunFifteenInjSLOneBetaZero &
        \RunFifteenInjSLOneInitialTimeZero & \RunFifteenInjSLOneAmplitudeDisplacementOne & \RunFifteenInjSLOneBetaOne & 
        \RunFifteenInjSLOneInitialTimeOne & 
        \xmark & \xmark & \xmark & \xmark & \xmark & \xmark\\
        15 & 
        \RunFifteenSLOneAmplitudeDisplacementZero & \RunFifteenSLOneBetaZero &
        \RunFifteenSLOneInitialTimeZero & \RunFifteenSLOneAmplitudeDisplacementOne & \RunFifteenSLOneBetaOne & 
        \RunFifteenSLOneInitialTimeOne & 
        \xmark & \xmark & \xmark & \xmark & \xmark & \xmark\\
    \hline
         &
        \RunTwentyEightInjSLOneAmplitudeDisplacementZero &
        \RunTwentyEightInjSLOneBetaZero &
        \RunTwentyEightInjSLOneInitialTimeZero &
        \RunTwentyEightInjSLTwoAmplitudeDisplacementZero &
        \RunTwentyEightInjSLTwoBetaZero &
        \RunTwentyEightInjSLTwoInitialTimeZero &
        \RunTwentyEightInjSLThreeAmplitudeDisplacementZero &
        \RunTwentyEightInjSLThreeBetaZero &
        \RunTwentyEightInjSLThreeInitialTimeZero &
        \RunTwentyEightInjSLFourAmplitudeDisplacementZero &
        \RunTwentyEightInjSLFourBetaZero &
        \RunTwentyEightInjSLFourInitialTimeZero \\
        28 &
        \RunTwentyEightSLOneAmplitudeDisplacementZero &
        \RunTwentyEightSLOneBetaZero &
        \RunTwentyEightSLOneInitialTimeZero &
        \RunTwentyEightSLTwoAmplitudeDisplacementZero &
        \RunTwentyEightSLTwoBetaZero &
        \RunTwentyEightSLTwoInitialTimeZero &
        \RunTwentyEightSLThreeAmplitudeDisplacementZero &
        \RunTwentyEightSLThreeBetaZero &
        \RunTwentyEightSLThreeInitialTimeZero &
        \RunTwentyEightSLFourAmplitudeDisplacementZero &
        \RunTwentyEightSLFourBetaZero &
        \RunTwentyEightSLFourInitialTimeZero \\
    \end{tabular}}
    \caption{
    Parameter estimation results assuming \textsc{Model D} glitches of increasing complexity.     White rows show the injected values and teal rows show the recovered median and 90\% confidence interval.  
    In particular, we consider a single-component glitch (ID 22), a glitch with two components (ID 15), and two glitches separated by 200 seconds with two components each (ID 28).
    The posterior distribution for the latter, most complex case is shown in Fig~\ref{fig:corner4D}.
    }\label{tab:glitchDisp_param}
\end{table*}

\subsection[Glitch only]{Inference with glitches alone, without GWs}
\label{subsec:glitchonly} 

\begin{figure*}[tbh!]
    \centering
    \includegraphics[width=0.45\textwidth, keepaspectratio]{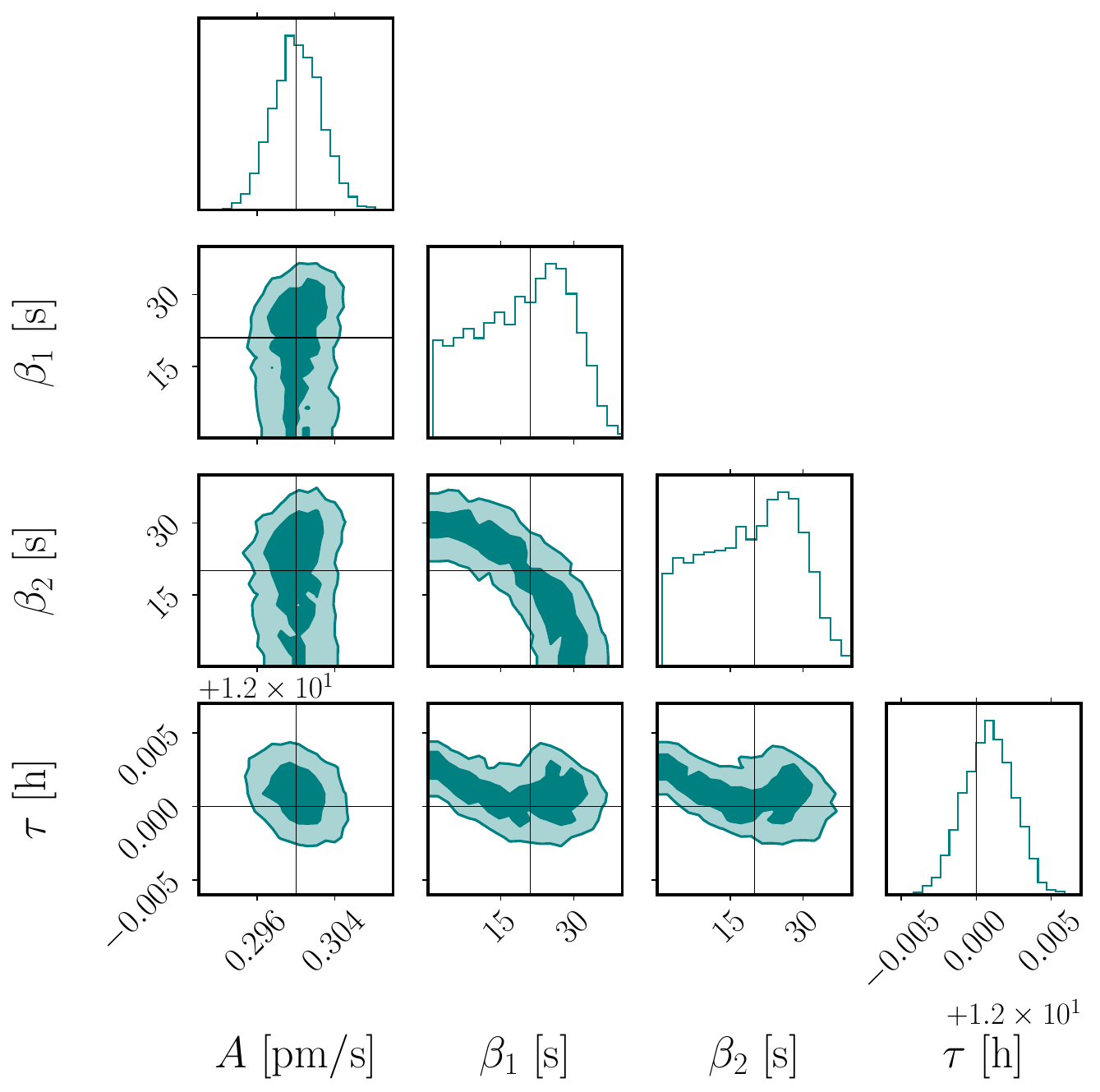}\\
    \vspace{.5cm}
    \includegraphics[width=0.45\textwidth, keepaspectratio]{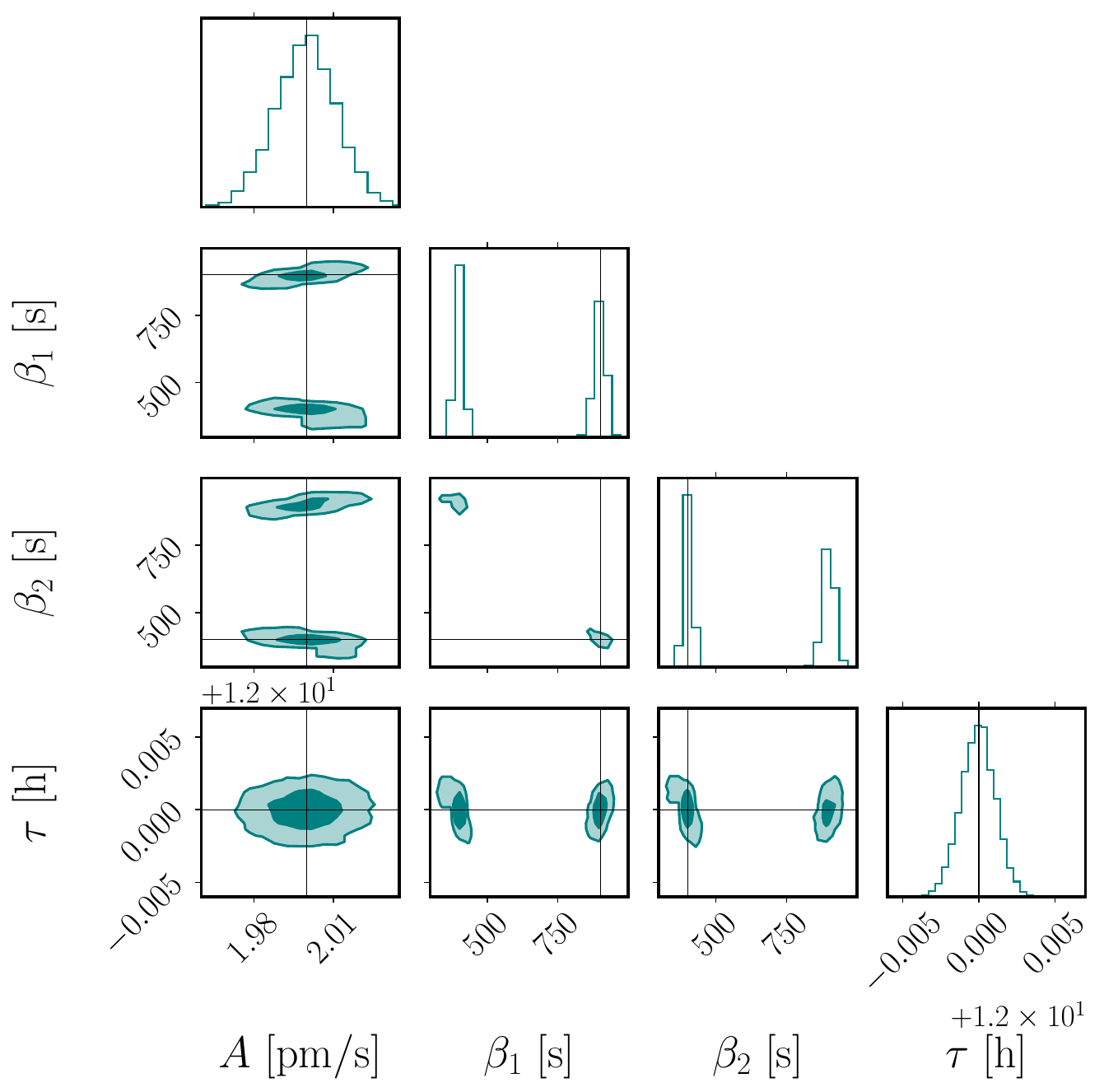}
    \hspace{.8cm}
    \includegraphics[width=0.45\textwidth, keepaspectratio]{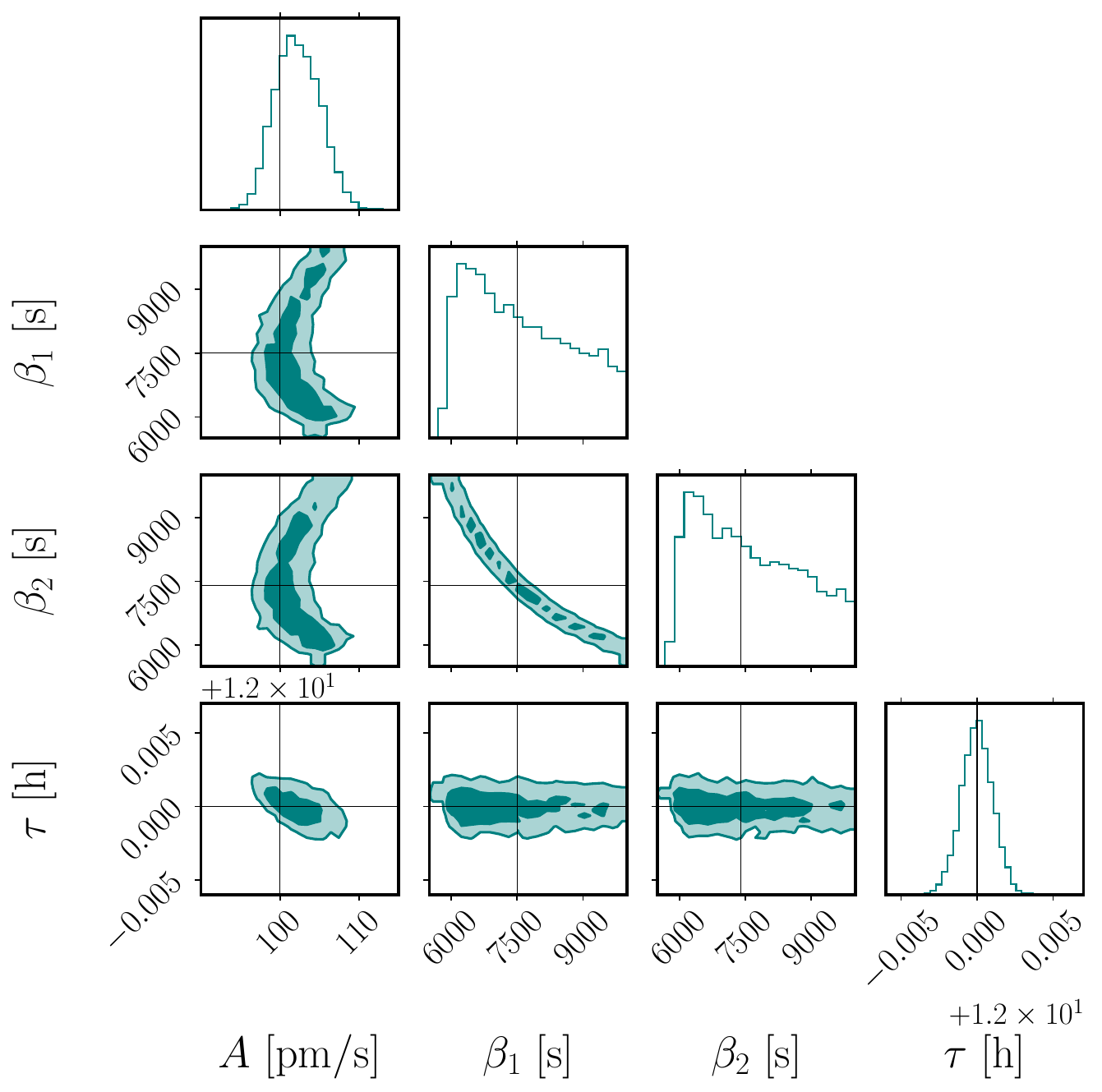}
    \caption{
   Posterior distributions for short (top, ID 30), medium (bottom left, ID 31), and long (bottom right, ID 32) \textsc{Model A1} glitches. Injected values and some posterior summary statistics are listed in Table~\ref{tab:glitchAcc_param}.     Darker (lighter) shaded areas indicate 90\% (50\%) credible regions and solid black lines indicate the injected values.
    The correlation between the fall time $\beta_1$ and the rise time $\beta_2$ is caused by the intrinsic degeneracy between these two parameters, see Eqs.~\eqref{eq:A1-TD} and~\eqref{eq:A1-FD}. For the medium-duration glitch, the larger separation between the injected value of $\beta_1-\beta_2 = 500 \mathrm{s}$ partially breaks it into a strong multimodality.
    }
    \label{fig:cornerLPFlmh}
\end{figure*}

\begin{figure*}[tbh!]
    \centering
    \includegraphics[width=0.65\textwidth, keepaspectratio]{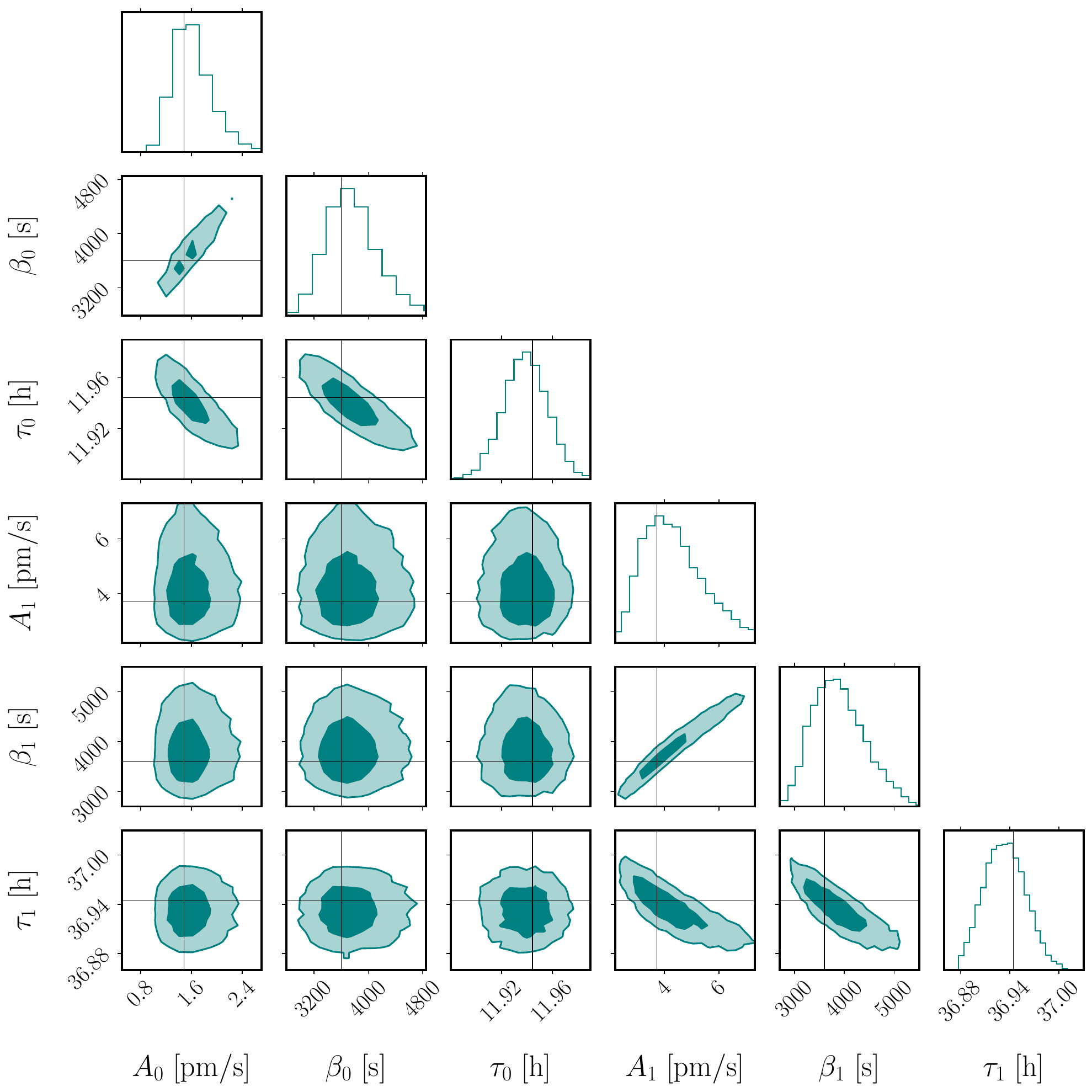}
    \caption{
    Posterior distributions for two \textsc{Model A2} glitches (run ID 29). 
    Injected values and some posterior summary statistics are listed in Table~\ref{tab:glitchAcc_param}.     Darker (lighter) shaded areas indicate 90\% (50\%) credible regions and solid black lines indicate the injected values.
The lower-left panels show the joint distribution between parameters describing the two glitches, which do not present significant correlations. }
    \label{fig:cornerA2}
\end{figure*}

\begin{figure*}[tbh!]
    \centering
    \includegraphics[width=\textwidth, keepaspectratio]{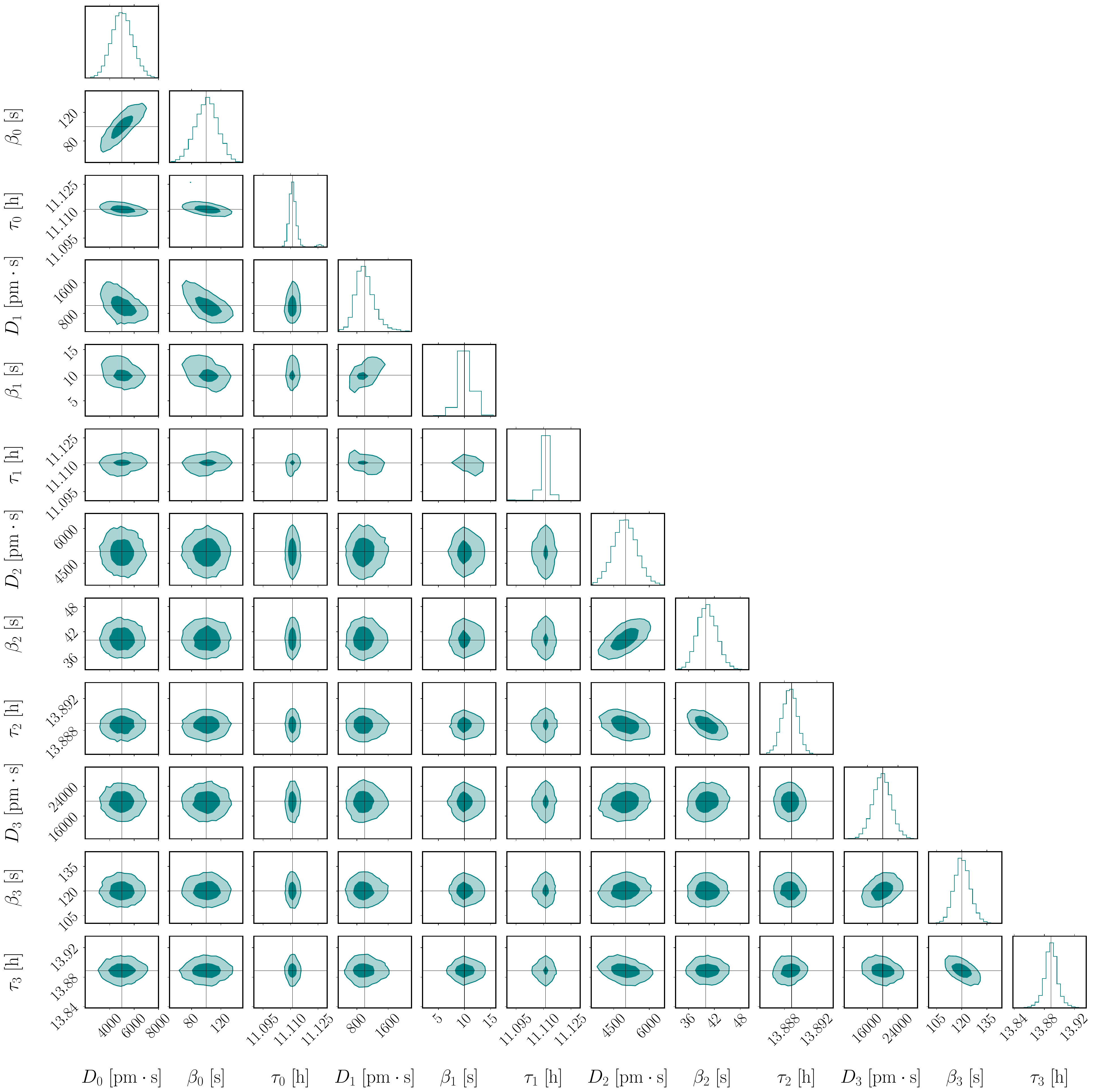}
    \caption{
     Posterior distributions for two \textsc{Model D} glitches (run ID 28). 
    Injected values and some posterior summary statistics are listed in Table~\ref{tab:glitchDisp_param}. 
      Each glitch is made of two components with injected values $\tau_0=\tau_1$ and $\tau_2=\tau_3$ for the first and second glitch, respectively. 
       Darker (lighter) shaded areas indicate 90\% (50\%) credible regions and solid black lines indicate the injected values.
    Glitch parameters are recovered successfully and cross-glitch correlations are negligible.}
    \label{fig:corner4D}
\end{figure*}

We consider all three glitch models presented in Sec.~\ref{sec:model} and inject them separately in the LISA data stream. Results are shown in Figs.~\ref{fig:cornerLPFlmh},~\ref{fig:cornerA2}, and~\ref{fig:corner4D} as well as Tables~\ref{tab:glitchAcc_param} and~\ref{tab:glitchDisp_param}. 

We perform model selection with different (i) number and order of shapelet components, (ii) number of glitches, and (iii) injection point. 
In particular, in Tab.~\ref{tab:summary-runs-glitchonly} we report ``strong'' evidence in favor of the correct noise-transient model for the selection of the number and order of shapelets; these are discrete parameters we can confidently identify using $\log_{10}\mathcal{B}_{15}^{j}$ with $j=16,\dots,20$.
We obtain a ``substantial" evidence $\log_{10}\mathcal{B}_{15}^{21}=0.9$ for selecting the correct number of glitches. Injection points are selected with a ``decisive'' evidence given by $\mathcal{B}_{22}^{n}$ with $n=23,\dots,27$.

All runs point to the same, encouraging result: glitch parameters are confidently reconstructed. 
In particular, we recover amplitudes across all models (i.e. $A$, $A_{0,1}$, $D_{0,1,2,3}$) with accuracies of $1\%-30\%$  at 90\% credible level.
Glitch-onset times are recovered with fractional accuracy $\lesssim 0.1\%$. 
The parameters $\beta_i$'s in \textsc{Model D} glitches are recovered with an accuracy of 20\%. 
On the other hand, \textsc{Model A1} glitches exhibit correlation and multimodalities for the joint posterior on $\beta_1$ and  $\beta_2$.
This is expected given the waveform degeneracy upon  exchange of these two parameters, cf. Eqs.~\eqref{eq:A1-TD} and Eq.~\eqref{eq:A1-FD}.

\section{\label{sec:conclusions} Conclusions}

We presented a parameter-estimation strategy to simultaneously extract GWs from MBHBs and glitches from future LISA data.
We developed several models for noise transients inspired by those observed by LPF. Crucially, we point out that dealing with glitches in the frequency domain greatly benefits from expressing the LISA response function (i.e. the TDIs) in terms of acceleration instead of displacement as usually done.

Accounting for potential noise transients in the data leads to accurate reconstruction of all GW parameters without significant correlations with the glitch properties. On the contrary, ignoring glitches when present in the data might introduce significant systematic biases on the reconstructed parameters of the MBHB.
Our analysis shows that  the most crucial property is the length of the glitch, with results ranging from a complete loss of the GW signal to a negligible impact.
When considering glitches in isolation, our procedure allows for confident identification of their number, location, and morphology in each of the models considered.

It is important to stress that all glitch models in our suite have a relatively low number of parameters and these are largely uncorrelated to those of the GW source. The computational overhead of including potential glitches in the signal model is therefore negligible, thus making our approach promising for a future ``global fit'' procedure.

This study is restricted to a single, fiducial GW source as well as glitches are conservatively placed at the time location that maximizes their matches with the GW signal. A broader injection-recovery study over the full MBHB  and glitch parameter space is needed to forecast the impact of noise transients on GW signals in the future LISA catalog; this is left to future work.

Overall, this paper showcases our readiness to model and precisely recover glitches when present in the LISA data stream, even when overlapping with GW sources of similar duration such as a MBHB.

\begin{acknowledgments}

We thank Chris Moore, Federico Pozzoli, Eleonora Castelli, Natalia Korsakova, Stas Babak, Martina Muratore, Nathan Steinle, and all \textsc{Balrog} developers for useful comments and inputs.
A.S. and D.G. are supported by ERC Starting Grant No. 945155--GWmining, Cariplo Foundation Grant No. 2021-0555, and MUR PRIN Grant No. 2022-Z9X4XS. 
A.S., D.G., and R.B. are supported by the ICSC National Research Center funded by NextGenerationEU. 
R.D., M.C., S.V., D.V.,W.J.W. acknowledge funding from MUR under the grant PRIN 2017-MB8AEZ.
R.B. acknowledges support through the Italian Space Agency grant \emph{Phase A activity for LISA mission, Agreement n. 2017-29-H.0}. 
D.G. is supported by Leverhulme Trust Grant No. RPG-2019-350. 
Computational work was performed using University of Birmingham BlueBEAR High Performance Computing facility and CINECA with allocations through INFN, Bicocca, and ISCRA project HP10BEQ9JB.

\textit{Software}:
We acknowledge usage of 
\textsc{Mathematica}~\cite{Mathematica} 
and of the following 
\textsc{Python}~\cite{10.5555/1593511} 
packages for modeling, analysis, post-processing, and production of results throughout:
\textsc{Nessai}~\cite{2021PhRvD.103j3006W},
\textsc{matplotlib}~\cite{2007CSE.....9...90H},
\textsc{numpy}~\cite{2020Natur.585..357H},
\textsc{scipy}~\cite{2020NatMe..17..261V}.
\end{acknowledgments}


\clearpage
\bibliography{biblio}
\clearpage
\onecolumngrid
\appendix

\end{document}